\documentclass[preprint,aps,fleqn]{revtex4}%
\usepackage{graphicx}
\usepackage{psfrag}
\usepackage{amsmath}
\usepackage{subfigure}

\begin{document}
\def\be{\begin{eqnarray}}
\def\ee{\end{eqnarray}}
\newcommand{\nn}{\nonumber}
\def\mpcomm#1{\nextline\strut\kern-6em{\tt MP COMMENT => \ #1}\nextline}
\def\nextline{\hfill\break}
\newcommand{\rf}[1]{(\ref{#1})}
\newcommand{\beq}{\begin{equation}}
\newcommand{\eeq}{\end{equation}}
\newcommand{\bea}{\begin{eqnarray}}
\newcommand{\eea}{\end{eqnarray}}
\newcommand{\pint}{-\hspace{-11pt}\int_{-\infty}^\infty }
\newcommand{\Pint}{-\hspace{-13pt}\int_{-\infty}^\infty }
\newcommand{\nint}{\int_{\infty}^{\infty}}
\renewcommand{\vec}[1]{\boldsymbol{#1}}

\title{Classical Strongly Coupled QGP: \\
VII. Shear Viscosity and Self Diffusion}
\author{Sungtae Cho}
\email{scho@grad.physics.sunysb.edu}
\author{Ismail Zahed}
\email{zahed@zahed.physics.sunysb.edu}
\affiliation{Department of Physics and Astronomy\\
State University of New York, Stony Brook, NY 11794-3800}

{\begin{abstract} We construct the Liouville operator for the
SU(2) classical colored Coulomb plasma (cQGP) for arbitrary values
of the Coulomb coupling $\Gamma=V/K$, the ratio of the mean
Coulomb to kinetic energy. We show that its resolvent in the
classical colored phase space obeys a hierarchy of equations. We
use a free streaming approximation to close the hierarchy and
derive an integral equation for the time-dependent structure
factor. Its reduction by projection yields hydrodynamical
equations in the long-wavelength limit.  We discuss the character
of the hydrodynamical modes at strong coupling. The shear
viscosity is shown to exhibit a minimum at $\Gamma\approx 8$ near
the liquid point. This minimum follows from the cross-over between
the single particle collisional regime which drops as
$1/\Gamma^{5/2}$ and the hydrodynamical collisional regime which
rises as $\Gamma^{1/2}$. The self-diffusion constant drops as
$1/\Gamma^{3/2}$ irrespective of the regime. We compare our
results to molecular dynamics simulations of the SU(2) colored
Coulomb plasma. We also discuss the relevance of our results for
the quantum and strongly coupled quark  gluon plasma (sQGP).
\end{abstract}
}

 \maketitle

\section{Introduction}

High temperature QCD is expected to asymptote a weakly coupled
Coulomb plasma albeit with still strong infrared divergences. The
latters cause its magnetic sector to be non-perturbative at all
temperatures. At intermediate temperatures of relevance to
heavy-ion collider experiments, the electric sector is believed to
be strongly coupled.

Recently, Shuryak and Zahed~\cite{SZ_newqgp} have suggested that
certain aspects of the quak-gluon plasma in range of temperatures
$(1-3)\,T_c$ can be understood by a stronger Coulomb interaction
causing persistent correlations in singlet and colored channels.
As a result the quark and gluon plasma is more a liquid than a gas
at intermediate temperatures. A liquid plasma should exhibit
shorter mean-free paths and stronger color dissipation, both of
which are supported by the current experiments at
RHIC~\cite{hydro}.

To help understand transport and dissipation in the strongly
coupled quark gluon plasma, a classical model of the colored
plasma was suggested in~\cite{gelmanetal}. The model consists of
massive quarks and gluons interacting via classical colored
Coulomb interactions. The color is assumed classical with all
equations of motion following from Poisson brackets. For the SU(2)
version both molecular dynamics simulations~\cite{gelmanetal} and
bulk thermodynamics~\cite{cho&zahed, cho&zahed2} were recently
presented including simulations of the energy loss of heavy
quarks~\cite{dusling&zahed}.

In this paper we extend our recent equilibrium analysis of the
static properties of the colored Coulomb plasma, to transport. In
section 2 we discuss the classical equations of motion in the
SU(2) colored phase space and derive the pertinent Liouville
operator. In section 3,  we show that the resolvent of the
Liouville operator obeys a hierarchy of equations in the SU(2)
phase space. In section 4 we derive an integral equation for the
time-dependent structure factor by introducing a non-local
self-energy  kernel in phase space. In section 5, we close the
Liouville hierarchy through a free streaming approximation on the
4-point resolvent and derive the self-energy kernel in closed form.
In section 6, we project the self-energy kernel and the non-static
structure factor onto the colorless hydrodynamical phase space.
 In section 7, we show that the sound and plasmon mode are the
 leading hydrodynamical modes in the SU(2) colored Coulomb plasma.
 In section we analyze the shear viscosity for the transverse sound mode
 for arbitrary values of $\Gamma$.  We show that a minimum forms at
$\Gamma\approx 5$ at the cross-over between the hydrodynamical
and single-particle regimes. In section 8, we analyze self-diffusion in
phase space, and derive an explicit expression for the diffusion
constant at strong coupling. Our conclusions and prospects are
in section 9.  In appendix A we briefly summarize our variables
in the SU(2) phase space. In appendix B we detail the projection
method for the self-energy kernel used in the text. In appendix C
we show that the collisional color contribution to the Liouville
operator drops in the self-energy kernel. In appendix D some useful
aspects of the hydrodynamical projection method are outlined.

\section{Colored Liouville Operator}

\renewcommand{\theequation}{II.\arabic{equation}}
\setcounter{equation}{0}

The canonical approach to the colored Coulomb plasma was discussed
in~\cite{gelmanetal}. In brief, the Hamiltonian for a single
species of constituent quarks or gluons in the SU(2)
representation is defined as

\begin{equation}
H=\sum_{i}^N\frac{\vec p^2_i}{2m_i}+\sum_{i>j=1}^N\frac{\vec
Q_i\cdot\vec Q_j}{|\vec r_i-\vec r_j|} \label{HAMILTON}
\end{equation}
The charge $g^2/4\pi$ has been omitted for simplicity of the
notation flow and will be reinserted in the pertinent physical
quantities by inspection.

The equations of motion in phase space follows from the classical
Poisson brackets. In particular

\begin{equation}
\frac{d\vec r_i}{dt}=-\{H,\vec r_i\}= \frac{\partial H}{\partial
\vec p_j}\frac{\partial \vec r_i}{\partial \vec r_j}=\frac{\vec
p_i}m \label{XPB}
\end{equation}
The Newtonian equation of motion is just the colored electric Lorentz
force

\begin{equation}
\frac{d\vec p_i}{dt}=-\{H,\vec p_i\}=- \frac{\partial H}{\partial
\vec r_j}\frac{\partial \vec p_i}{\partial \vec p_j}= Q_i^a\,\vec
E_i^a=\vec{F}_i \label{PPB}
\end{equation}
with the colored electric field and potentials defined as
($a=1,2,3$)

\begin{equation}
\vec{E}_i^a=-\vec\nabla_i\Phi^a_i=-\vec\nabla_i\sum_{j\neq
i}\frac{Q^a_j}{|\vec r_i-\vec r_j|} \label{XPB2}
\end{equation}
Our strongly coupled colored plasma is mostly electric following
the original assumptions in~\cite{gelmanetal,gelmanetal2}. The
equation of motion of the color charges is

\begin{equation}
\frac{dQ_i^a}{dt}=-\{H,Q_i^a\}= -\sum_{j,b}\frac{\partial
H}{\partial Q_i^b}\frac{\partial Q_i^a}{\partial Q_j^c} \{Q_j^b,
Q_j^c\}=\sum_{j\neq i}\frac{Q_iT^aQ_j}{|\vec r_i-\vec r_j|}
\label{QPB}
\end{equation}
for arbitrary color representation. For SU(2) the classical color
charge (\ref{QPB}) precesses around the net colored potential
$\Phi$ determined by the other particles as defined in
(\ref{XPB2}),

\begin{equation}
\frac{d\vec Q_i}{dt}=(\vec \Phi_i\times \vec Q_i)
\end{equation}
This equation was initially derived by Wong~\cite{wong}.
Some aspects of the SU(2) phase space are briefly recalled
in Appendix A.

The set (\ref{XPB}), (\ref{PPB}) and (\ref{QPB}) define the
canonical evolution in phase space. The time-dependent phase
distribution is formally given by

\begin{equation}
f(t,\vec r \vec p \vec Q)=\sum_{i=1}^N\delta(\vec r-\vec
r_i(t))\delta(\vec p-\vec p_i(t))\delta(\vec Q-\vec Q_i(t))\equiv
\sum_i\delta(\vec q-\vec q_i(t)) \label{FDIS}
\end{equation}
For simplicity $\vec q$ is generic for $\vec r,\vec p,\vec Q$.
Using the chain rule, the time-evolution operator on (\ref{FDIS})
obeys

\begin{equation}
\frac{d}{dt}=\frac{\partial}{\partial t}+ \frac{d\vec
r_i}{dt}\frac{\partial}{\partial \vec r_i}+ \frac{d\vec
p_i}{dt}\frac{\partial}{\partial \vec p_i}+ \frac{d\vec
Q_i}{dt}\frac{\partial}{\partial \vec Q_i} \equiv
\partial_t+i\mathcal{L} \label{LIOU}
\end{equation}
The last relation defines the Liouville operator

\begin{equation}
\mathcal{L}=\mathcal{L}_0+\mathcal{L}_I+\mathcal{L}_Q=-i\vec
v_i\cdot\nabla_{\vec r_i}-i\,\vec{F}_i\cdot\vec{\nabla}_{\vec p_i}
-i\vec \Phi_i \cdot( \vec Q_i\times\vec \nabla_{\vec Q_i})
\label{LIOU1}
\end{equation}
The last contribution in (\ref{LIOU1}) is genuily a 3-body force
because of the cross product (orbital color operator). It requires
3 distinct colors to not vanish. This observation will be
important in simplifying the color dynamics below.  Also
(\ref{LIOU1}) is hermitean.

Since (\ref{FDIS}) depends implicitly on time, we can write
formally

\begin{equation}
\frac{d}{dt}f(t,\vec r\vec p\vec Q)=i\mathcal{L}f(t,\vec r\vec
p\vec Q) \label{EVOLUTION}
\end{equation}
with a solution $f(t)=e^{i\mathcal{L}t}f(0)$. The formal relation
(\ref{EVOLUTION}) should be considered with care since the action
of the Liouville operator on the 1-body phase space distribution
(\ref{FDIS}) generates also a 2-body phase space distribution.
Indeed, while $\mathcal{L}_0$ is local in phase space

\begin{equation}
\mathcal{L}_0\sum_i\delta(\vec q-\vec q_i)=-i\vec
v\cdot\nabla_{\vec r}\sum_i\delta(\vec q-\vec q_i)=L_0(\vec
q)\sum_i\delta(\vec q-\vec q_i) \label{LO}
\end{equation}
the 2 other contributions are not. Specifically

\begin{eqnarray}
\mathcal{L}_I\sum_m\delta(\vec q-\vec q_m)&=&i\sum_{i\neq
j}\nabla_{\vec r_i} \frac{\vec Q_i\cdot\vec Q_j}{|\vec r_i-\vec
r_j|}
\cdot \nabla_{\vec p_i}\sum_m \delta(\vec q-\vec q_m) \nonumber \\
&=& i\int d\vec q'   \sum_{i\neq j , mn}\nabla_{\vec
r_i}\frac{\vec Q_i\cdot\vec Q_j}{|\vec r_i-\vec
r_j|}\cdot\nabla_{\vec p_i}\,\,\delta(\vec
q-\vec q_m) \delta(\vec q'-\vec q_n) \nonumber \\
&=&- \int d\vec q'L_I(\vec q,\vec q')\,\sum_{mn}\delta(\vec q-\vec
q_m)\delta(\vec q'-\vec q_n)
\label{L1}
\end{eqnarray}
with

\begin{equation}
L_I(\vec q,\vec q')=i\nabla_{\vec r} \frac{\vec Q\cdot\vec
Q'}{|\vec r-\vec r'|} \cdot (\nabla_{\vec p}- \nabla_{\vec p'})
\end{equation}
Similarly

\begin{eqnarray}
\mathcal{L}_Q\sum_m\delta(\vec q-\vec q_m)&=&-i\sum_{ j\neq i, m}
\frac{\vec Q_i\times\vec Q_j}{|\vec r_i-\vec r_j|}\cdot
\nabla_{\vec Q_i} \,\delta(\vec q-\vec q_m) \nonumber \\
&=& -i\int d\vec q'\sum_{j\neq i, mn}\frac{\vec Q_i\times \vec
Q_j}{|\vec r_i-\vec r_j|}\cdot \nabla_{\vec Q_i}\delta(\vec
q-\vec q_m) \delta(\vec q'-\vec q_n) \nonumber \\
&=&- \int d\vec q'L_Q(\vec q,\vec q') \sum_{mn}\delta(\vec q-\vec
q_m)\delta(\vec q'-\vec q_n) \label{LQ}
\end{eqnarray}
with

\begin{equation}
L_Q(\vec q,\vec q')=-i \frac{\vec Q\times\vec Q'}{|\vec r-\vec
r'|} \cdot (\nabla_{\vec Q}-\nabla_{\vec Q'})
\end{equation}
Clearly (\ref{LQ}) drops from 2-body and symmetric phase space
distributions. It does not for 3-body and higher.

\section{Liouville Hierarchy}

\renewcommand{\theequation}{III.\arabic{equation}}
\setcounter{equation}{0}

An important correlation function in the analysis of the colored
Coulomb plasma is the time dependent structure factor or 2-body
correlation in the color phase space

\begin{equation}
{\bf S}(t-t',\vec r-\vec r', \vec p\vec p', \vec Q\cdot \vec
Q')=\langle\delta f(t,\vec r\vec p\vec Q)\,\delta f(t',\vec r'\vec
p'\vec Q')\rangle \label{FF}
\end{equation}
with $\delta f=f-\langle f\rangle$ the shifted 1-body phase space
distribution.  The averaging in (\ref{FF}) is carried over the
initial conditions with fixed number of particles $N$ and average
energy or temperature $\beta=1/T$. Thus $\langle f\rangle=nf_0(p)$
which is the Maxwellian distribution for constituent quarks or
gluons. In equilibrium, the averaging in (\ref{FF}) is time and
space  translational invariant as well as color rotational
invariant.

Using the ket notation with $\vec 1\equiv \vec q\equiv {\vec
r}{\vec p}{\vec Q}$

\begin{equation}
|\delta f(t,\vec 1)\rangle=\bigg|\sum_{m}\delta(\vec q-\vec
q_m(t))-\langle \sum_{m}\delta(\vec q-\vec q_m(t))\bigg\rangle
\equiv |\delta f(t,\vec 1)-\langle f(t,\vec 1)\rangle>
\end{equation}
with also $\vec 2=\vec q'$, $\vec 3=\vec q''$, $\vec 4=\vec q'''$
and so on and the formal Liouville solution $\delta f(t,\vec
1)=e^{i\mathcal{L}t}\delta f(\vec 1)$ we can write (\ref{FF}) as

\begin{equation}
{\bf S}(t-t',\vec q,\vec q')=\langle\delta f(t,\vec 1)|\delta
f(t',\vec 2)\rangle= \langle\delta f(\vec 1)|e^{i{\cal
L}(t'-t)}|\delta f(\vec 2)\rangle
\end{equation}
The bra-ket notation is short for the initial or equilibrium
average. Its Laplace or causal transform reads

\begin{equation}
{\bf S}(z, \vec q, \vec q')=-i\int_{-\infty}^{+\infty} \,dt
\,\,\theta(t-t')\,\, e^{izt} \,\,{\bf S}(t-t',\vec q,\vec q')=
\langle\delta f(\vec 1)|\frac 1{z+\mathcal{L}}|\delta f(\vec
2)\rangle \label{RES}
\end{equation}
with $z=\omega+i0$. Clearly

\begin{equation}
z{\bf S}(z, \vec q, \vec q')+\langle\delta f(\vec
1)|\mathcal{L}\frac 1{z+\mathcal{L} }|\delta f(\vec
2)\rangle=\langle\delta f(\vec 1)|\delta f(\vec 2)\rangle
\label{RES1}
\end{equation}
Since ${\cal L }^\dagger={\cal L}$ is hermitian and using
(\ref{LO}), (\ref{L1}) and (\ref{LQ}) it follows that

\begin{equation}
\langle\delta f(\vec 1)|{\cal L}=\langle\delta f(\vec 1)|{
L}_0(\vec q)- \int \,d\vec q"\,L_{I+Q}(\vec q,\vec
q'')\,\langle\delta f(\vec 1)\delta f(\vec 3)|
\end{equation}
Thus

\begin{equation}
\Big(z-L_0(\vec q)\Big){\bf S}(z, \vec q,\vec q')-\int\,d\vec
q''L_{I+Q}(\vec q',\vec q''){\bf S}(z,\vec q\vec q'',\vec q')={\bf
S}_0(\vec q,\vec q') \label{RES2}
\end{equation}
where we have defined the 3-body phase space resolvent

\begin{equation}
{\bf S}(z, \vec q\vec q'', \vec q')=\,\langle\delta f(\vec
1)\delta f(\vec 3)|\frac 1{z+{\cal L}}\,|\delta f(\vec 2)>
\label{SS3}
\end{equation}
${\bf S}_0(\vec q,\vec q')$ is the static colored structure factor
discussed by us in~\cite{cho&zahed3}. Since $L_{I+Q}(\vec q',\vec
q)$ is odd under the switch $\vec q\leftrightarrow \vec q'$, and
since ${\bf S}(z, \vec q\vec q'',\vec q')={\bf S}(-z, \vec q,\vec
q'\vec q'')$ owing to the $t\leftrightarrow t'$ in (\ref{RES}),
then

\begin{equation}
\Big(z+L_0(\vec q')\Big){\bf S}(z, \vec q,\vec q')-\int\,d\vec
q''L_{I+Q}(\vec q',\vec q''){\bf S}(z,\vec q, \vec q' \vec
q'')={\bf S}_0(\vec q,\vec q') \label{RES2X}
\end{equation}
(\ref{RES2}) or equivalently (\ref{RES2X}) define the Liouville
hierarchy, whereby the 2-body phase space distribution ties to the
3-body phase space distribution and so on.  Indeed,  (\ref{RES2X})
for instance implies

\begin{equation}
\Big(z+L_0(\vec q'')\Big){\bf S}(z, \vec q\vec q', \vec
q'')-\int\,d\vec q'''L_{I+Q}(\vec q'',\vec q'''){\bf S}(z,\vec q
\vec q', \vec q''\vec q''')={\bf S}_0(\vec q\vec q',\vec q'')
\label{RES2XX}
\end{equation}
with the 4-point resolvent function

\begin{equation}
{\bf S}(z,\vec q \vec q', \vec q''\vec q''')= \,\langle\delta
f(\vec 1)\delta f(\vec 2)|\frac 1{z+{\cal L}}\,|\delta f(\vec
3)\delta f(\vec 4)> \label{SS3X}
\end{equation}
These are the microscopic kinetic equations for the color phase
space distributions.  They are only useful when closed, that is by
a truncation as we discuss below. These formal equations where
initially discussed
in~\cite{foster&martin,foster,mazenko1,mazenko3} in the context of
the one component Coulomb Abelian Coulomb plasma. We have now
generalized them to the multi-component and non-Abelian colored
Coulomb plasma.

\section{Self-Energy Kernel}

\renewcommand{\theequation}{IV.\arabic{equation}}
\setcounter{equation}{0}

In (\ref{RES2}) the non-local part of the Liouville operator plays
the role of a non-local self-energy kernel $\Sigma$ on the 2-body
resolvent.  Indeed, we can rewrite (\ref{RES2}) as

\begin{eqnarray}
(z-L_0(\vec q)){\bf S}(z,\vec q,\vec q')-\int dq''\Sigma (z,\vec
q,\vec q'') {\bf S}(z,\vec q'',\vec q')={\bf S}_0(\vec q,\vec q')
\label{SELF}
\end{eqnarray}
with the non-local self-energy kernel defined formally as

\begin{equation}
\int d\vec q''\Sigma(z,\vec q,\vec q'')\,{\bf S}(z,\vec q'',\vec
q')=\int d\vec q'' L_{I+Q}(\vec q,\vec q'')\,{\bf S}(z,\vec q\vec
q'',\vec q') \label{SELF1}
\end{equation}
The self-energy kernel $\Sigma$ can be regarded as the sum of a
static or z-independent contribution $\Sigma_S$ ans a non-static
or collisional contribution $\Sigma_C$,

\begin{equation}
\Sigma(z, \vec q, \vec q'')=\Sigma_S(\vec q,\vec
q'')+\Sigma_C(z,\vec q,\vec q'') \label{SUM}
\end{equation}
The stationary part $\Sigma_S$ satisfies

\begin{equation}
\int d\vec q'' \Sigma_S(\vec q,\vec q'')\,{\bf S}_0(\vec q'',\vec
q')= \int d\vec q''L_{I+Q}(\vec q,\vec q'')\,{\bf S}_0(\vec q,\vec
q',\vec q'') \label{SELF2}
\end{equation}
which identifies it with the sum of the 2- and 3-body part of the
Liouville operator $L_{I+Q}$.

The collisional part $\Sigma_C$ is more involved. To unwind it, we
operate with $(z+L_0(\vec q'))$ on both sides of (\ref{SELF1}),
and then reduce the left hand side contribution using
(\ref{RES2X}) and the right hand side contribution using
(\ref{RES2XX}).  The outcome reduces to

\begin{eqnarray}
\Sigma_C(z, \vec q, \vec q'')\,{\bf S}_0(\vec q'',\vec q')=&&
-\int d\vec q'''\,L_{I+Q}(\vec q,\vec q'')L_{I+Q}(\vec q',\vec
q'')
\,{\bf S}(z,\vec q\vec q'',\vec q'\vec q''')\nonumber\\
&&+\int\,dq'''\,\Sigma(z,\vec q,\vec q'')\,L_{I+Q}(\vec q',\vec
q'')\,{\bf S}(z, \vec q'', \vec q'\vec q''') \label{INT}
\end{eqnarray}
after using (\ref{SELF2}).  From (\ref{SELF1}) it follows formally
that

\begin{equation}
\Sigma(z,\vec q,\vec q'')=\int\,d\vec q'''\, L_{I+Q}(\vec q,\vec
q''')\,{\bf S}^{-1}(z,\vec q',\vec q'')\,{\bf S}(z, \vec q\vec
q''',\vec q') \label{FORMX}
\end{equation}
Inserting (\ref{FORMX}) into the right hand side of (\ref{INT})
and taking the $\vec q'$ integration on both sides yield

\begin{equation}
nf_0(p'')\,\Sigma_C(z,\vec q,\vec q'') =-\int d\vec q'' d\vec
q'''\,L_{I+Q}(\vec q,\vec q'')L_{I+Q} (\vec q',\vec q''')\,{\bf
G}(z,\vec q\vec q'',\vec q'\vec q''') \label{SELF5}
\end{equation}
with ${\bf G}$ a 4-point phase space correlation function

\begin{equation}
{\bf G}(z,\vec q\vec q_1',\vec q'\vec q_2')={\bf S}(z,\vec q\vec
q_1',\vec q'\vec q_2')-\int d\vec q_3d\vec q_4 {\bf S}(z, \vec
q\vec q_1',\vec q_3){\bf S}^{-1}(z,\vec q_3,\vec q_4){\bf
S}(z,\vec q_4,\vec q'\vec q_2') \label{G4}
\end{equation}
The collisional character of the self-energy $\Sigma_C$ is
manifest in (\ref{SELF5}).  The formal relation for the
collisional self-energy (\ref{SELF5}) was initially derived
in~\cite{mazenko1,mazenko3} for the one-component  and Abelian
Coulomb plasma. We now have shown that it holds for any
non-Abelian SU(N) Coulomb plasma.

Eq. (\ref{SELF5}) shows that the connected part of the self-energy
kernel is actually tied to a 4-point correlator in the colored
phase space. In terms of (\ref{SELF5}), the original kinetic
equation (\ref{RES2}) now reads

\begin{eqnarray}
&&\Big(z-L_0(\vec q)\Big){\bf S}(z, \vec q, \vec q')-\int\,d\vec
q''\Sigma_S(\vec q,\vec q''){\bf S}(z,\vec q'',\vec q')= {\bf
S}_0(\vec q,\vec q') \nonumber\\&& -\int\,d\vec q''\,d\vec q_1
d\vec q_2\,L_{I+Q}(\vec q,\vec q_1)L_{I+Q} (\vec q'',\vec
q_2)\,{\bf G}(z,\vec q\vec q_1,\vec q''\vec q_2)\,{\bf S}(z,\vec
q'',\vec q') \label{BOLZ}
\end{eqnarray}
which is a Boltzman-like equation. The key difference is that it
involves correlation functions and the Boltzman-like kernel in the
right-hand side is {\bf not} a scattering amplitude but rather a
reduced 4-point correlation function.  (\ref{BOLZ}) reduces to the
Boltzman equation for weak coupling. An alternative derivation of
(\ref{BOLZ}) can be found in Appendix C through a direct
projection of (\ref{SELF1}) in phase space.

\section{Free Streaming Approximation}

\renewcommand{\theequation}{V.\arabic{equation}}
\setcounter{equation}{0}

The formal kinetic equation (\ref{SELF5})  can be closed by
approximating the 4-point correlation function in the color phase
space by a product of 2-point correlation
function~\cite{mazenko3},

\begin{equation}
{\bf G}(t, \vec q\vec q_1, \vec q'\vec q_2)\approx \Big( {\bf
S}(t, \vec q,\vec q'){\bf S}(t, \vec q_1,\vec q_2)+{\bf S}(t, \vec
q, \vec q_2){\bf S}(t, \vec q', \vec q_1) \Big)\label{ST3}
\end{equation}
This reduction will be referred to as the free steaming
approximation. Next we substitue the colored Coulomb potentials in
the double Liouville operator $L_{1+Q}\times L_{1+Q}$ with a bare
Coulomb ${\bf V}(\vec r-\vec r',\vec Q\cdot\vec Q')=\vec
Q\cdot\vec Q'/|\vec r-\vec r'|$.

\begin{eqnarray}
& & L_{I+Q}(\vec q,\vec q_1)=i\nabla_{\vec r}{\bf V}(\vec r-\vec
r_1,\vec Q\cdot\vec
Q_1)\cdot (\nabla_{\vec p}-\nabla_{\vec p_1}) \nonumber \\
& &-i \Big(\vec Q\times\nabla_{\vec Q}{\bf V}(\vec r-\vec r_1,\vec
Q\cdot \vec Q_1)\cdot \nabla_{\vec Q}+\vec Q_1\times \nabla_{\vec
Q_1}{\bf V}(\vec r-\vec r_1,\vec Q\cdot \vec
Q_1)\cdot\nabla_{Q_1}\Big) \label{ST4}
\end{eqnarray}
times a dressed colored Coulomb potential ${\bf c}_D$ defined
in~\cite{cho&zahed3}

\begin{eqnarray}
& & L_{I+Q}^R(\vec q,\vec q_1)=-i\frac{1}{\beta}\nabla_{\vec
r}{\bf c}_D(\vec r-\vec r_1,\vec Q\cdot \vec Q_1)\cdot
(\nabla_{\vec p}-\nabla_{\vec p_1}) \nonumber \\
& & +i\frac{1}{\beta}\Big(\vec Q\times\nabla_{\vec Q}{\bf
c}_D(\vec r-\vec r_1,\vec Q\cdot \vec Q_1)\cdot \nabla_{\vec
Q}+\vec Q_1\times \nabla_{\vec Q_1}{\bf c}_D(\vec r-\vec r_1,\vec
Q\cdot \vec Q_1)\cdot\nabla_{Q_1}\Big) \label{ST5}
\end{eqnarray}
This bare-dressed or half renormalization was initially
suggested~\cite{wallenborn&baus} in the context of the
one-component Coulomb plasma to overcome the shortcomings of a
full or dressed-dressed renormalization initially suggested
in~\cite{mazenko1,mazenko3}. The latter was shown to upset the
initial conditions. Thus

\begin{equation}
L_{I+Q}(\vec q,\vec q_1)L_{I+Q}(\vec q',\vec q_2)\rightarrow
\frac{1}{2}\Big( L_{I+Q}(\vec q,\vec q_1)L_{I+Q}^R(\vec q',\vec
q_2)+L_{I+Q}^R(\vec q,\vec q_1)L_{I+Q}(\vec q',\vec q_2)\Big)
\label{eq004s}
\end{equation}
Combining (\ref{ST3}) and (\ref{eq004s}) in (\ref{SELF5}) yields


\begin{eqnarray}
&& n\,f_0(\vec p')\,\Sigma_C(t, \vec q, \vec q')\approx
-\frac{1}{2}\int d\vec q_1\,d\vec q_2\,\bigg(L_{I+Q}(\vec q,\vec
q_1)L_{I+Q}^R (\vec q',\vec q_2){\bf S}(t, \vec q,\vec q'){\bf
S}(t,\vec q_1,\vec q_2) \nonumber\\
&& +L_{I+Q}(\vec q,\vec q_1)L_{I+Q}^R (\vec q',\vec q_2){\bf S}(t,
\vec q,\vec q_2){\bf S}(t,\vec q', \vec q_1) + (\vec
q_1\leftrightarrow \vec q_2,\vec q\leftrightarrow \vec q' ) \bigg)
\label{ST6}
\end{eqnarray}



\noindent This is the half dressed but free streaming
approximation for the connected part of the self-energy for the
colored Coulomb plasma. Translational invariance in space and
rotational invariance in color space allows a further reduction of
(\ref{ST6}) by Fourier and Legendre transforms respectively.
Indeed, Eq. (\ref{ST6}) yields

\begin{eqnarray}
&&n\,f_0(\vec p')\,\Sigma_C(t, \vec q, \vec
q')\nonumber\\
&&\approx -\frac{1}{2} \int d\vec q_1\,d\vec q_2\,\bigg(L_{I}(\vec
q,\vec q_1)L_{I}^R (\vec q',\vec q_2){\bf S}(t, \vec q,\vec
q'){\bf S}(t,\vec q_1,\vec q_2) \nonumber\\
&& +L_{I}(\vec q,\vec q_1)L_{I}^R (\vec q',\vec q_2){\bf S}(t,
\vec q,\vec q_2){\bf S}(t,\vec q', \vec q_1) + (\vec
q_1\leftrightarrow \vec q_2,\vec q\leftrightarrow \vec q') \bigg) \nonumber \\
&&=-\frac{1}{2\beta} \int d\vec q_1\,d\vec q_2\,\bigg(
\nabla_{\vec r}{\bf c}_D(\vec r-\vec r_1,\vec Q\cdot \vec
Q_1)\cdot \nabla_{\vec p}\nabla_{\vec r'}{\bf V}(\vec r'-\vec
r_2,\vec Q'\cdot \vec Q_2)\cdot \nabla_{\vec
p'}{\bf S}(t, \vec q,\vec q'){\bf S}(t,\vec q_1,\vec q_2) \nonumber\\
&& +\nabla_{\vec r}{\bf c}_D(\vec r-\vec r_1,\vec Q\cdot \vec
Q_1)\cdot \nabla_{\vec p}\nabla_{\vec r'}{\bf V}(\vec r'-\vec
r_2,\vec Q'\cdot \vec Q_2)\cdot \nabla_{\vec p'}{\bf S}(t, \vec
q,\vec q_2){\bf S}(t,\vec q', \vec q_1) + (\vec q_1\leftrightarrow
\vec q_2,\vec q\leftrightarrow \vec q') \bigg) \nonumber \\
\label{eq006s}
\end{eqnarray}
where we note that the colored part of the Liouville operator
dropped from the collision kernel in the free streaming
approximation as we detail in Appendix C. Both sides of
(\ref{eq006s}) can be now Legendre transformed in color to give

\begin{eqnarray}
&& n\,f_0(\vec p')\,\sum_l\Sigma_{Cl}(t, \vec r\vec r', \vec p\vec
p')\frac{2l+1}{4\pi}P_l(\vec Q\cdot\vec Q') \nonumber \\
& & \approx -\frac 1{2\beta}\int d\vec r_1 d\vec p_1 d\vec
r_2d\vec p_2 \sum_{l}\frac{2l+1}{4\pi}\bigg(\frac{l+1}
{2l+1}P_{l+1}(\vec Q\cdot\vec Q')+\frac{l}{2l+1}P_{l-1}
(\vec Q\cdot\vec Q')\bigg) \nonumber \\
& & \times \bigg( \nabla_{\vec r}{\bf c}_{D1}(\vec r-\vec
r_1)\cdot \nabla_{\vec p}\nabla_{\vec r'} \frac{1}{|\vec r'-\vec
r_2|}\cdot \nabla_{\vec p'}{\bf S}_l(t, \vec r\vec r',\vec p\vec
p'){\bf S}_1(t, \vec r_1\vec r_2,\vec p_1\vec p_2) \nonumber \\
& & +\nabla_{\vec r}{\bf c}_{Dl}(\vec r-\vec r_1)\cdot
\nabla_{\vec p}\nabla_{\vec r'} \frac{1}{|\vec r'-\vec r_2|}\cdot
\nabla_{\vec p'}{\bf S}_1(t, \vec r\vec r_2,\vec p\vec p_2){\bf
S}_l(t, \vec r'\vec r_1,\vec p'\vec p_1) \nonumber \\
& & +\nabla_{\vec r'}{\bf c}_{Dl}(\vec r'-\vec r_2)\cdot
\nabla_{\vec p'}\nabla_{\vec r} \frac{1}{|\vec r-\vec r_1|}\cdot
\nabla_{\vec p}{\bf S}_l(t, \vec r\vec r_2,\vec p\vec p_2){\bf
S}_1(t, \vec r'\vec r_1,\vec p'\vec p_1) \nonumber \\
& & \nabla_{\vec r'}{\bf c}_{D1}(\vec r'-\vec r_2)\cdot
\nabla_{\vec p'}\nabla_{\vec r} \frac{1}{|\vec r-\vec r_1|}\cdot
\nabla_{\vec p}{\bf S}_l(t, \vec r\vec r',\vec p\vec p'){\bf
S}_1(t, \vec r_1\vec r_2,\vec p_1\vec p_2) \bigg) \label{eq007s}
\end{eqnarray}
Thus

\begin{eqnarray}
&& n\,f_0(\vec p')\,\Sigma_{Cl}(t, \vec r\vec r', \vec p\vec
p') \nonumber \\
& & \approx -\frac 1{2\beta}\int d\vec r_1 d\vec p_1 d\vec
r_2d\vec p_2 \bigg( \nabla_{\vec r}{\bf c}_{D1}(\vec r-\vec
r_1)\cdot \nabla_{\vec p}\nabla_{\vec r'} \frac{1}{|\vec
r'-\vec r_2|}\cdot \nabla_{\vec p'} \nonumber \\
& & \times \Big(\frac{l}{2l+1}{\bf S}_{l-1}(t, \vec r\vec r',\vec
p\vec p'){\bf S}_1(t, \vec r_1\vec r_2,\vec p_1\vec
p_2)+\frac{l+1}{2l+1}{\bf S}_{l+1}(t, \vec r\vec r',\vec p\vec
p'){\bf S}_1(t, \vec r_1\vec r_2,\vec p_1\vec
p_2)\Big) \nonumber \\
& & +\nabla_{\vec r}{\bf c}_{D1}(\vec r-\vec r_1)\cdot
\nabla_{\vec p}\nabla_{\vec r'} \frac{1}{|\vec r'-\vec r_2|}\cdot
\nabla_{\vec p'} \nonumber \\
& & \times \Big(\frac{l}{2l+1}{\bf S}_{1}(t, \vec r\vec r_2,\vec
p\vec p_2){\bf S}_{l-1}(t, \vec r'\vec r_1,\vec p'\vec p_1)+\frac{
l+1}{2l+1}{\bf S}_{1}(t, \vec r\vec r_2,\vec p\vec p_2){\bf
S}_{l+1}(t, \vec r'\vec r_1,\vec p'\vec p_1)\Big) \nonumber \\
& & +\nabla_{\vec r'}{\bf c}_{Dl}(\vec r'-\vec r_2)\cdot
\nabla_{\vec p'}\nabla_{\vec r} \frac{1}{|\vec r-\vec r_1|}\cdot
\nabla_{\vec p} \nonumber \\
& & \times \Big(\frac{l}{2l+1}{\bf S}_{l-1}(t, \vec r\vec r_2,\vec
p\vec p_2){\bf S}_1(t, \vec r'\vec r_1,\vec p'\vec p_1)+\frac{l+1}
{2l+1}{\bf S}_{l+1}(t, \vec r\vec r_2,\vec p\vec p_2){\bf
S}_1(t, \vec r'\vec r_1,\vec p'\vec p_1) \Big)  \nonumber \\
& & +\nabla_{\vec r'}{\bf c}_{D1}(\vec r'-\vec r_2)\cdot
\nabla_{\vec p'}\nabla_{\vec r} \frac{1}{|\vec r-\vec r_1|}\cdot
\nabla_{\vec p} \nonumber \\
& & \times \Big(\frac{l}{2l+1}{\bf S}_{l-1}(t, \vec r\vec r',\vec
p\vec p'){\bf S}_1(t, \vec r_1\vec r_2,\vec p_1\vec
p_2)+\frac{l+1}{2l+1}{\bf S}_{l+1}(t, \vec r\vec r',\vec p\vec
p'){\bf S}_1(t, \vec r_1\vec r_2,\vec p_1\vec p_2)\Big)  \bigg)
\nonumber \\ \label{eq008s}
\end{eqnarray}
with ${\bf S}_{l-1}\equiv0$ by definition. In the colored Coulomb
plasma the collisional contributions diagonalize in the color
projected channels labelled by $l$, with $l=0$ being the density
channel, $l=1$ the plasmon channel and so on.  In momentum space
(\ref{eq008s}) reads

\begin{eqnarray}
&& n\,f_0(\vec p')\, \Sigma_{Cl}(t, \vec k, \vec p\vec p')
\nonumber \\
& & = -\frac 1{2\beta} \int d\vec p_1 d\vec p_2 \int \frac{d\vec
l}{(2\pi)^3}\bigg( \vec l\cdot\nabla_{\vec p}\vec
l\cdot\nabla_{\vec p'} {\bf c}_{D1}(l) V_{\vec l} \nonumber
\\ & & \times \Big( \frac{l}{2l+1} {\bf S}_{l-1}(t, \vec k-\vec
l,\vec p\vec p'){\bf S}_1(t, \vec l,\vec p_1\vec
p_2)+\frac{l+1}{2l+1} {\bf S}_{l+1}(t, \vec k-\vec l,\vec p\vec
p'){\bf S}_1(t, \vec l,\vec p_1\vec p_2)\Big) \nonumber \\
& & + \vec l\cdot\nabla_{\vec p}(\vec k-\vec l)\cdot\nabla_{\vec
p'} {\bf c}_{Dl}(l) V_{\vec k-\vec l} \nonumber
\\ & & \times \Big( \frac{l}{2l+1} {\bf S}_{1}(t, \vec k-\vec
l,\vec p\vec p_2){\bf S}_{l-1}(t, \vec l,\vec p'\vec
p_1)+\frac{l+1}{2l+1} {\bf S}_{1}(t, \vec k-\vec l,\vec p\vec
p_2){\bf S}_{1+1}(t, \vec l,\vec p'\vec p_1)\Big) \nonumber \\
& & + (\vec k-\vec l)\cdot\nabla_{\vec p}\vec l\cdot\nabla_{\vec
p'} {\bf c}_{Dl}(l) V_{\vec k-\vec l} \nonumber
\\ & & \times \Big( \frac{l}{2l+1} {\bf S}_{l-1}(t, \vec
l,\vec p\vec p_2){\bf S}_{l}(t, \vec k-\vec l,\vec p'\vec
p_1)+\frac{l+1}{2l+1} {\bf S}_{l+1}(t, \vec l,\vec p\vec
p_2){\bf S}_{1}(t, \vec k-\vec l,\vec p'\vec p_1)\Big) \nonumber \\
& & +\vec l\cdot\nabla_{\vec p}\vec l\cdot\nabla_{\vec p'} {\bf
c}_{D1}(l) V_{\vec l} \nonumber \\
& & \times \Big( \frac{l}{2l+1} {\bf S}_{l-1}(t, \vec k-\vec
l,\vec p\vec p'){\bf S}_1(t, \vec l,\vec p_1\vec
p_2)+\frac{l+1}{2l+1} {\bf S}_{l+1}(t, \vec k-\vec l,\vec p\vec
p'){\bf S}_1(t, \vec l,\vec p_1\vec p_2)\Big) \bigg) \nonumber \\
\label{SFOURIER}
\end{eqnarray}
with $V_{\vec l}=4\pi/{\vec l}^2$. We note that for $l=0$ which is
the colorless density channel (\ref{SFOURIER}) involves only ${\bf
S}_1$ which is the time-dependent charged form factor due to the
Coulomb interactions.

\section{Hydrodynamical Projection}

\renewcommand{\theequation}{VI.\arabic{equation}}
\setcounter{equation}{0}

In terms of (\ref{SFOURIER}) , (\ref{SELF1}) and

\begin{equation}
\Sigma_l (z\vec k, \vec p\vec p_1)=\bigg(
\Sigma_{0l}+\Sigma_{Il}+\Sigma_{Ql}+\Sigma_{Cl}\bigg)(z\vec k,
\vec p\vec p_1) \label{KEY}
\end{equation}
the Fourier and Legendre transform of the kinetic equation
(\ref{RES2})  now read

\begin{equation}
z{\bf S}_l(z\vec k, \vec p\vec p')-\int\,d\vec p_1\Sigma_l(z\vec
k, \vec p\vec p_1){\bf S}_l(z\vec k, \vec p_1\vec p')={\bf
S}_{0l}(\vec k, \vec p\vec p') \label{K1}
\end{equation}
with $\Sigma_{0l}=L_0$ and $\Sigma_{Sl}=L_{(I+Q)l}$. Specifically

\begin{eqnarray}
&& \Sigma_{0l}(z\vec k,\vec p\vec p_1)=\vec k\cdot\vec v \delta(\vec p -\vec p_1)\nonumber\\
&& \Sigma_{Il}(z\vec k,\vec p\vec p_1)=-n\,f_0(p)\frac{\vec k\cdot\vec p}{m}\,{\bf c}_{Dl}(\vec k)\nonumber\\
&& \Sigma_{Ql}(z\vec k,\vec p\vec p_1)=0 \label{K2}
\end{eqnarray}
and $\Sigma_{Cl}$ is defined in (\ref{SFOURIER}). See also
Appendix B for an alternative but equivalent derivation using the
operator projection method.

(\ref{K1}) is the key kinetic equation for the colored Coulomb
plasma. It still contains considerable information in phase space.
A special limit of the classical phase space is the long
wavelength or hydrodynamical limit. In this limit, only few
moments of the phase space fluctuations $\delta f$ or equivalently
their correlations in ${\bf S}\approx \langle\delta f\delta
f\rangle$ will be of interest. In particular,

\begin{eqnarray}
&& \vec{n}(t, \vec r)=\int d\vec{p} d\vec Q\,
\, \delta f (t, \vec r, \vec p, \vec Q)\nonumber\\
&& \vec{p}(t, \vec r)=\int d\vec{p} d\vec Q\, \vec p
\, \delta f (t, \vec r, \vec p, \vec Q)\nonumber\\
&& {\bf e}(t, \vec r)=\int d\vec{p} d\vec Q \,\frac {p^2}{2m}\,
\delta f (t, \vec r, \vec p, \vec Q) \label{MOM}
\end{eqnarray}
The local particle density, 3-momentum and energy (kinetic).  The
hydrodynamical sector described by the macro-variables (\ref{MOM})
is colorless. An interesting macro-variable which carries charge
representation of SU(2) would be

\begin{equation}
{\bf n}_l (t, \vec r)=\frac 1{2l+1} \sum_m \int d\vec{r} d\vec Q
\,Y_l^m (\vec Q)\, \delta f (t, \vec r, \vec p, \vec Q)
\label{DENSL}
\end{equation}
which reduces to the $l$ color density with $l=0$ being the
particle density, $l=1$ the charged color monopole density,  $l=2$
the charged color quadrupole density and so on.  Because of color
rotational invariance in the SU(2) colored Coulomb plasma, the
constitutive equations for (\ref{DENSL}) which amount to charge
conservation hold for each $l$.

To project (\ref{K1}) onto the hydrodynamical part of the phase
space characterized by (\ref{DENSL}) and (\ref{MOM}), we define
the hydrodynamical projectors

\begin{equation}
{\cal P}_H=\sum_{i=1}^5|i\rangle\langle i|\qquad\qquad {\cal
Q}_H={\bf 1}_H-{\cal P}_H \label{PRO}
\end{equation}
with $1=$ l-density, $2,4,5=$ momentum and $3=$ energy as detailed
in Appendix D. When the $l=0$ particle density is retained in
(\ref{PRO}) the projection is on the colorless sector of the phase
space. When the $l=1$ charged monopole density is retained in
(\ref{PRO}) the projection is on the plasmon channel, and so on.
Most of the discussion to follow will focus on projecting on the
canonical hydrodynamical phase space (\ref{MOM}) with $l=0$ or
singlet representation. The inclusion of the $l\neq 0$
representations of SU(2)  is straightforward.

Formally (\ref{KEY}) can be viewed as a $\vec p\times \vec p_1$
matrix in momentum space

\begin{equation}
\left( z-\Sigma_l(z\vec k)\right)\,{\bf S}_l(z\vec k)={\bf
S}_{0l}(\vec k) \label{KEY1}
\end{equation}
The projection of the matrix equation (\ref{KEY1}) follows the
same procedure as in Appendix B. The result is

\begin{equation}
\left(z-{\cal P}_H\Sigma_l (z\vec k){\cal P}_H-{\cal
P}_H\Theta_l(z\vec k){\cal P}_H\right) {\cal P}_H{\bf S}_l(z\vec
k){\cal P}_H={\cal P}_H{\bf S}_{0l}(k){\cal P}_H \label{KEY2}
\end{equation}
with

\begin{equation}
\Theta_l=\Sigma_l(z\vec k) {\cal Q}_H(z-{\cal Q}_H\Sigma_H(z\vec
k){\cal Q}_H)^{-1} {\cal Q}_H\Sigma_l(z\vec k)
\end{equation}
If we define the hydrodynamical matrix elements

\begin{eqnarray}
&&{\bf G}_{lij}(z\vec k)=\langle i|{\bf S}_l(z\vec k)|j\rangle\nonumber\\
&&{\Sigma}_{lij}(z\vec k)=\langle i|{\Sigma }_l(z\vec k)|j\rangle\nonumber\\
&&{\Theta}_{lij}(z\vec k)=\langle i|{\Theta}_l(z\vec k)|j\rangle\nonumber\\
&&{\bf G}_{0lij}(z\vec k)=\langle i|{\bf S}_{0l}(\vec k)|j\rangle
\label{MAT}
\end{eqnarray}
then (\ref{KEY2}) reads

\begin{equation}
\left(z\delta_{ii'}-\Omega_{lij}(z\vec k)\right)\,{\bf
G}_{lji'}(z\vec k)={\bf G}_{0lii'}(\vec k) \label{DIS}
\end{equation}
with $\Omega_l=\Sigma_l+\Theta_l$. (\ref{DIS}) takes the form of a
dispersion for each color partial wave $l$ with the projection
operator (\ref{PRO}) set by the pertinent density (\ref{DENSL}).
The contribution $\Sigma_l$ to $\Omega_l$ will be referred to as
{\it direct} while the contribution $\Theta_l$ will be referred to
as {\it indirect}.

\section{Hydrodynamical Modes}

\renewcommand{\theequation}{VII.\arabic{equation}}
\setcounter{equation}{0}

The zeros of (\ref{DIS}) are the hydrodynamical modes originating
from the Liouville equation for the time-dependent structure factor. The
equation is closed under the free streaming approximation with
half renormalized vertices as we detailed above.

We start by analyzing the 2 transverse modes with $i=T$ in
(\ref{MAT}) and (\ref{DIS}). We note with~\cite{baus} that ${\bf
G}_{lTi}=0$ whenever $T\neq i$. The hydrodynamical projection (see
Appendix D) causes the integrand to be odd whatever $l$. The 2
independent transverse modes in (\ref{DIS}) decouple from the
longitudinal $i=L$, the (kinetic) energy $i=E$ and particle
density $i=N$ modes for all color projections. Thus

\begin{equation}
{\bf G}_{lT}(z\vec k)=\frac 1{z-\Omega_{lT}(z\vec k)} \label{TDIS}
\end{equation}
with $\Omega_{lT}=\langle T|\Omega_l|T\rangle$ and ${\bf
G}_{lT}=\langle T|{\bf G}_l|T\rangle$. The hydro-projected
time-dependent $l$ structure factor for fixed frequency
$z=\omega+i0$, wavenumber $k$ develops 2 transverse poles

\begin{equation}
z_l(\vec k)=\Omega_{lT}(z\vec k)\approx {\cal O}(k^2)
\label{TPOLE}
\end{equation}
The last estimate follows from O(3) momentum symmetry under
statistical averaging whatever the color projection.  We identify
the transverse poles in  (\ref{TPOLE}) with 2 shear modes of
consititutive dispersion

\begin{equation}
\omega+i\frac{\eta_l}{mn}k^2+{\cal O}(k^3)= 0
\label{VIS2}
\end{equation}
with $\eta_l$ the shear viscosity for the lth color
representation. Unlike conventional plasmas, the classical SU(2)
color Coulomb plasma admits an infinite hierarchy of shear modes
for each representation $l$.

\begin{figure}[!h]
\begin{center}
\subfigure{\label{radial:all}\includegraphics[width=0.495\textwidth]
{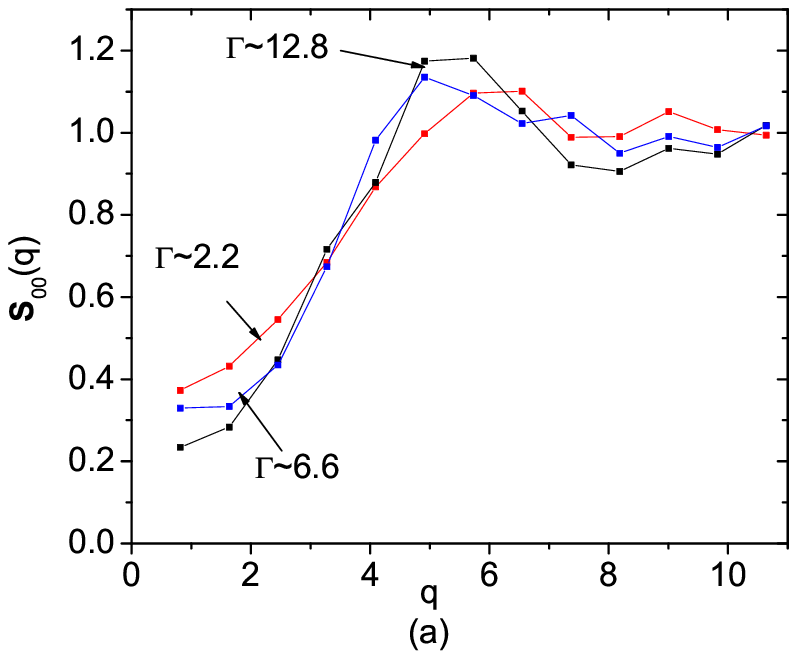}}
\subfigure{\label{radial:G021}\includegraphics[width=0.495\textwidth]
{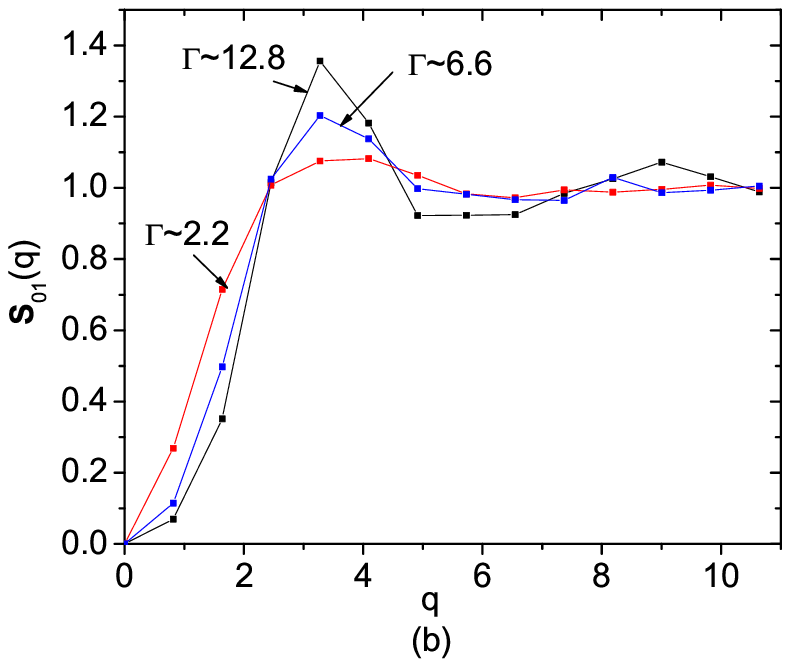}}
\end{center}
\begin{center}
\subfigure{\label{radial:G066}\includegraphics[width=0.495\textwidth]
{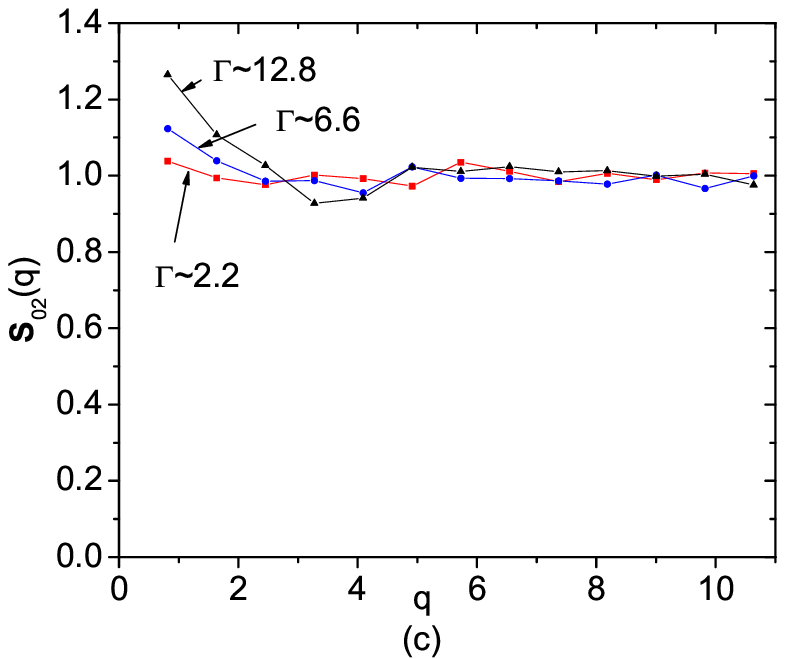}}
\subfigure{\label{radial:G128}\includegraphics[width=0.495\textwidth]
{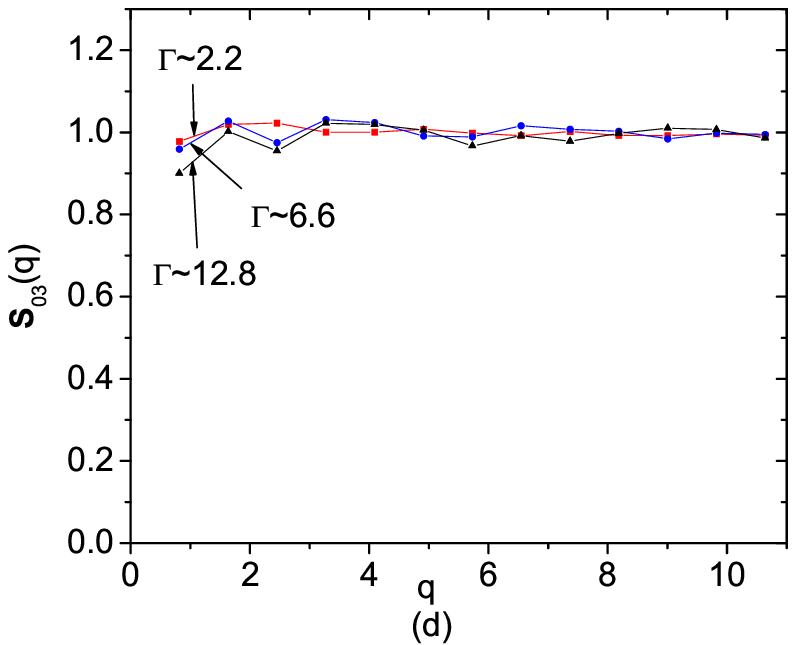}}
\end{center}
\caption{${\bf S}_{0l}(q)$ from SU(2) Molecular Dynamics.} \label{structure}
\end{figure}

\begin{figure}[!h]
\begin{center}
\includegraphics[width=0.55\textwidth]{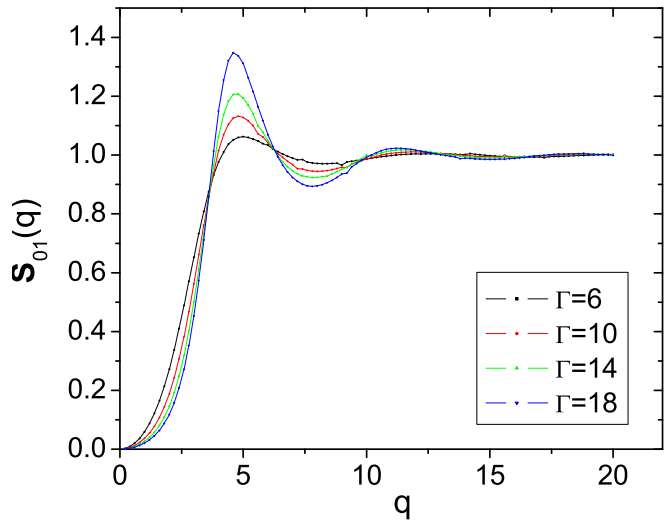}
\end{center}
\caption{${\bf S}_{01}(q)$ for different $\Gamma$~\cite{cho&zahed3}}.
\label{fluctuation}
\end{figure}

The remaining 3 hydrodynamical modes $L,E,N$ are more involved as
they mix in (\ref{DIS}) and under general symmetry consideration.
Indeed current conservation, ties the L mode to the N mode for
instance.  Most of the symmetry arguments regarding the generic
nature of $\Omega_l$ in~\cite{baus} carry to our case for each
color representation. Thus,  for the 3 remaining non-transverse
modes (\ref{DIS}) reads in matrix form

\begin{equation}
\left(
\begin{array}{ccc}
{\bf G}_{l NN} &  {\bf G}_{lNL} & {\bf G}_{lNE} \\
{\bf G}_{lLN}  &{\bf G}_{lLL}   & {\bf G}_{lLE}  \\
{\bf G}_{lEN}&  {\bf G}_{lEL} &   {\bf G}_{lEE}
\end{array}
\right)= \left(
\begin{array}{ccc}
z  &  -\Omega_{lNL} & 0  \\
-\Omega_{lLN}  &z-\Omega_{lLL}   & -\Omega_{lLE}  \\
0  &  -\Omega_{lEL} &   z-\Omega_{lEE}
\end{array}
\right)^{-1} \left(
\begin{array}{ccc}
1+n\,{\bf h}_l &  0 & 0  \\
0  &1 & 0  \\
0  & 0 &  1
\end{array}
\right) \label{MAT1}
\end{equation}
The 3 remaining hydrodynamical modes are the zeros of the
determinant

\begin{equation}
\Delta_l=\left|
\begin{array}{ccc}
z  &  -\Omega_{lNL} (zk)& 0  \\
-\Omega_{lLN} (zk) &z-\Omega_{lLL} (zk)  & -\Omega_{lLE} (zk) \\
0  &  -\Omega_{lEL} (zk) &   z-\Omega_{lEE} (zk)
\end{array}
\right|=0 \label{MAT2}
\end{equation}
(\ref{MAT2}) admits infinitly many solutions $z_l(k)$.  We seek
the hydrodynamical solutions as analytical solutions in $k$ for
small $k$, ie. $z(k)=\sum_nz_{ln}k^n$ for each SU(2) color
representation $l$.  In leading order, we have

\begin{equation}
\Delta_l\approx z_{l0}\left(z_{l0}^2-\frac{k^2T}{m}\,{\bf
S}_{0l}^{-1}(k\approx 0)\right)\approx 0 \label{MAT3}
\end{equation}
after using the symmetry properties of $\Omega_l$ as
in~\cite{baus} for each $l$.  We have also made use of the
generalized Ornstein-Zernicke equations for each $l$
representation~\cite{cho&zahed3}
In Fig.~\ref{structure} we show the molecular dynamics simulation
results for 4 typical structure factors~\cite{cho&zahed3}

\begin{equation}
{\bf S}_{0l}(\vec k )=\left({\frac{4\pi}{2l+1}}\right)
\left<\left| \sum_{jm}\,e^{{i\vec k}\cdot{\vec
x_j(0)}}\,Y_l^m(\vec Q_i )\,\right|^2 \right>
\end{equation}
for $l=0,1,2,3$. We have made use of the dimensionless wavenumber
$q=k\,a_{WS}$ with $a_{WS}$ is the Wigner-size radius.  In
Fig.~\ref{fluctuation} we show the analytical result for ${\bf
S}_{01}$ which we will use for the numerical estimates below.  We
note that the $l=1$ structure factor which amounts to the monopole
structure factor vanishes at $k=0$. All other $l$'s are finite at
$k=0$ with $l=0$ corresponding to the density structure factor.

(\ref{MAT3}) displays 3 hydrodynamical zeros as $k\rightarrow 0$
for each $l$ representation. One is massless and we identify it
with the diffusive heat mode. The molecular dynamics simulations
of the structure factors in Fig.~\ref{structure} implies that all
$l\neq 0$ channels are sound dominated with two massless
modes, while the $l=1$ is plasmon dominated with two massive
longitudinal plasmon states. Thus

\begin{equation}
z_{l\pm}=\pm \omega_{p}^2\delta_{l1}
\end{equation}
with $\omega_{p}=k_D\sqrt{T/m}$ the plasmon frequency. The
relevance of this channel to the energy loss has been discussed in
\cite{cho&zahed5}. We used ${\bf S}_{01}(k\approx 0)\approx
k^2/k_D^2$ with $k_D^2$ the squared Debye momentum. All even
$l\neq 1$ are contaminated by the sound modes. The SU(2) classical
and colored Coulomb plasma supports plasmon oscillations even at
strong coupling. These modes are important in the attenuation of
soft monopole color oscilations.

\section{Shear Viscosity}

\renewcommand{\theequation}{VIII.\arabic{equation}}
\setcounter{equation}{0}

The transport parameters associated to the SU(2) classical and colored Coulomb plasma
follows from the hydrodynamical projection and expansion discussed above. This includes,
the heat diffusion coefficient, the transverse shear viscosity and the longitudinal plasmon
frequency and damping parameters. In this section, we discuss explicitly the shear viscosity
coefficient for the SU(2) colored Coulomb plasma.

Throughout,  we define  $\lambda=\frac{4}{3}\pi(3\Gamma)^{3/2}$,
the bare Coulomb interaction  $\bar{V}_l=k_0^2/l^2$ in units of the Wigner-size radius $k^{-1}_0=a_{
WS}$.  While varying the Coulomb coupling

\begin{equation}
\Gamma=\frac{g^2}{4\pi}\beta\frac{C_2}{a_{WS}}
\end{equation}
 all length scales will be measured in  $a_{WS}=(4\pi n/3)^{-1/3}$,
all times in the inverse plasmon frequency $1/\omega_p$ with
$\omega_p^2={\kappa_D^2}/{m\beta}={n}g^2C_2/m$.
All units of mass will be measured in $m$. The Debye
momentum is $\kappa_D^2=g^2n\beta C_2$ and the plasma density is $n$.
 for instance, the shear viscosity will be expressed in fixed dimensionless units
of $\eta_*=nm\omega_p a_{WS}^{2}$.

The transverse shear viscosity follows from (\ref{TDIS}) with $\Sigma_l$ contributing to the direct
or hydrodynamical part, and $\Theta_l$ contributing to the indirect or single-particle part. For $l=0$

\begin{equation}
\frac{\eta_{0}}{\eta^*}=\frac{\eta_{0\,
dir}}{\eta^*}+\frac{\eta_{0\, ind}}{\eta^*} \label{ETAFULL}
\end{equation}
respectively. The direct or hydrodynamical
contribution is likely to be dominant at strong coupling, while
the indirect or single-particle contribution is likely to take
over at weak coupling. We now  proceed to show that.

The indirect contribution to the viscosity follows from the
contribution outside the hydrodynamical subspace through
${\cal Q}_H$ and lumps the single-particle phase contributions.
It  involves the inversion of  ${\cal Q}_H\Sigma_{C0}{\cal Q}_H$
in (\ref{eq013e}) with

\begin{eqnarray}
\eta_{0\rm ind}=\lim_{k\to0}\frac {mn}{k^2} \frac{|\langle
t|\Sigma_0 |tl\rangle|^2}{\langle tl|i\Sigma_0|tl\rangle}
=\lim_{k\to0}\frac {mn}{k^2} \frac{|\langle
t|(\Sigma_{00}+\Sigma_{C0}) |tl\rangle|^2}{\langle
tl|i\Sigma_{C0}|tl\rangle} \label{VIS3}
\end{eqnarray}
In short we expand $\Sigma_{C0}$ in terms of generalized Hermite
polynomials, with the first term identified with the stress tensor
due to the projection operator (\ref{eq003h}). The inversion
follows by means of the first Sonine polynomial expansion.
Explicitly

\begin{equation}
\eta_{ind}=\frac{\eta_{0\,ind}}{\eta_*}=nm \lim_{k \to
0}\frac{1}{k^2}\frac{ |\langle t \vert\Sigma_{00}+\Sigma_{C0}(\vec
k, 0)\vert lt \rangle |^2 }{\langle lt \vert i\Sigma_{C0}(\vec k,
0)\vert lt \rangle} =\frac{(1+\lambda I_2)^2} {\lambda I_3}
\label{eq013v}
\end{equation}
with

\begin{eqnarray}
& & I_2=\frac{1}{60\pi^2}\frac{1}{(3\Gamma)^{1/2}}
\int_0^{\infty} dq \Big(2({\bf S}_{01}(q)^2-1)+(1-{\bf S}_{01}(q))\Big) \nonumber \\
& & I_3=\frac{1}{10\pi^{3/2}}\frac{1}{3\Gamma} \int_0^{\infty}dq q
(1-{\bf S}_{01}(q)) \label{eq014v}
\end{eqnarray}
with the dimensionless wave number $q=ka_{WS}$.

We recall that ${\bf S}_{01}$ is the monopole structure factor
discussed in~\cite{cho&zahed3} both analytically and numerically.
In Fig.~\ref{fluctuation} we show the behavior of the static
monopole structure factor from~\cite{cho&zahed3} for different
Coulomb couplings. The larger $\Gamma$ the stronger the first
peak, and the oscillations. These features characterize the onset
of the crystalline structure in the SU(2) colored Coulomb plasma.
A good fit to Fig.~\ref{fluctuation}  follows from the following
parametrization

\begin{equation}
1+ C_0 e^{-q/C_1}\sin{((q-C_2)/C_3)} \label{FIT}
\end{equation}
with 4 parameters $C_{0,1,2,3}$.  The fit following from
(\ref{FIT}) extends to $q\approx 100$  within $10^{-5}$ accuracy,
thanks to the exponent.

The direct contribution to the shear viscosity follows from similar arguments.
From (\ref{TDIS}) and (\ref{VIS2}), we have in the zero momentum limit

\begin{eqnarray}
\eta_{0\, dir}=\lim_{k\to0} \frac {mn}{k^2}\langle t|i\Sigma_0
|t\rangle=\lim_{k\to0}\frac {mn}{k^2}\langle t|i\Sigma_{C0}
(0,0)|t\rangle
\end{eqnarray}
with $\Sigma_0=\Sigma_{00}+\Sigma_{I0}+\Sigma_{C0}$ as defined in
(\ref{K2}) and (\ref{SFOURIER}). Only those nonvanishing
contributions after the hydrodynamical projection were retained in
the second equalities in (\ref{VIS3})  as we detail in Appendix D.
A rerun of the arguments yields

\begin{eqnarray}
\eta_{dir}^{*}&=&\eta_{0\,dir}/\eta_*=\lambda\frac{\omega_p}
{\kappa_D^3}\lim_{k\to0}\frac{1}{\vec k^2}\int\frac{d\vec
l}{(2\pi)^3} \int_{0}^{\infty}dt
n(\vec \epsilon\cdot\vec l)^2  \nonumber \\
& \times & \bigg( {\bf c}_{D1}(l){\cal G}_{n1}(\vec k-\vec
l,t){\cal G}_{n1}(\vec l,t)\bar{V}_{\vec l}-{\bf c}_{D0}(l){\cal
G}_{n1}(\vec k-\vec l,t){\cal G}_{n1}(\vec l,t)\bar{V}_{\vec
k-\vec
l} \bigg) \nonumber \\
\label{eq002v}
\end{eqnarray}
The projected non-static structure factor is

\begin{eqnarray}
{\cal G}_{n1}(\vec l,t) &=&\frac{1}{n}\int d\vec pd\vec p'\,{\bf
S}_{1}(\vec l,t;\vec p\vec p')={\overline{\cal G}}_{n1}(\vec l,
t)\,{\bf S}_{01}(\vec l)  \label{eq010v}
\end{eqnarray}
with the normalization ${\overline{\cal G}}_{n1}(\vec l, 0)=1$. As
in the one component Coulomb plasma studied
in~\cite{gould&mazenko} we will approximate the dynamical part by
its intermediate time-behavior where the motion is free. This
consists in solving (\ref{SELF}) with no self-energy kernel or
$\Sigma=0$,

\begin{equation}
{\cal G}_{n1}(\vec l, t)\approx e^{-(lt)^2/2m\beta}\,{\bf
S}_{01}(\vec l) \label{GFREE}
\end{equation}
Thus inserting (\ref{GFREE}) and performing the integrations with
$k\rightarrow 0$ yield the direct contribution to the shear
viscosity

\begin{equation}
\eta_{dir}^{*}
=\frac{\eta_{dir}}{\eta_0}=\frac{\sqrt{3}}{45\pi^{1/2}}\Gamma^{\frac{1}{2}}
\label{eq012v}
\end{equation}

The full shear viscosity result is then
\begin{equation}
\frac{\eta_{0}}{\eta^*}=\frac{\eta_{0\,
dir}}{\eta^*}+\frac{\eta_{0\, ind}}{\eta^*}=
\frac{\sqrt{3}}{45\pi^{1/2}}\Gamma^{\frac{1}{2}} +\frac{(1+\lambda
I_2)^2} {\lambda I_3} \label{ETAFULLX}
\end{equation}
after inserting (\ref{eq013v}) and (\ref{eq012v}) in
(\ref{ETAFULL}). The result (\ref{ETAFULLX}) for the shear
viscosity of the transverse sound mode is analogous to the result
for the sound velocity in the one component plasma derived
initially in~\cite{wallenborn&baus} with two differences: 1/ The
SU(2) Casimir in $\Gamma$; 2/ the occurrence of ${\bf S}_{01}$
instead of ${\bf S}_{00}$.  Since ${\bf S}_{01}$ is plasmon
dominated at low momentum, we conclude that the shear viscosity is
dominated by rescattering against the SU(2) plasmon modes in the
cQGP.

\begin{figure}[!h]
\begin{center}
\includegraphics[width=0.49\textwidth]{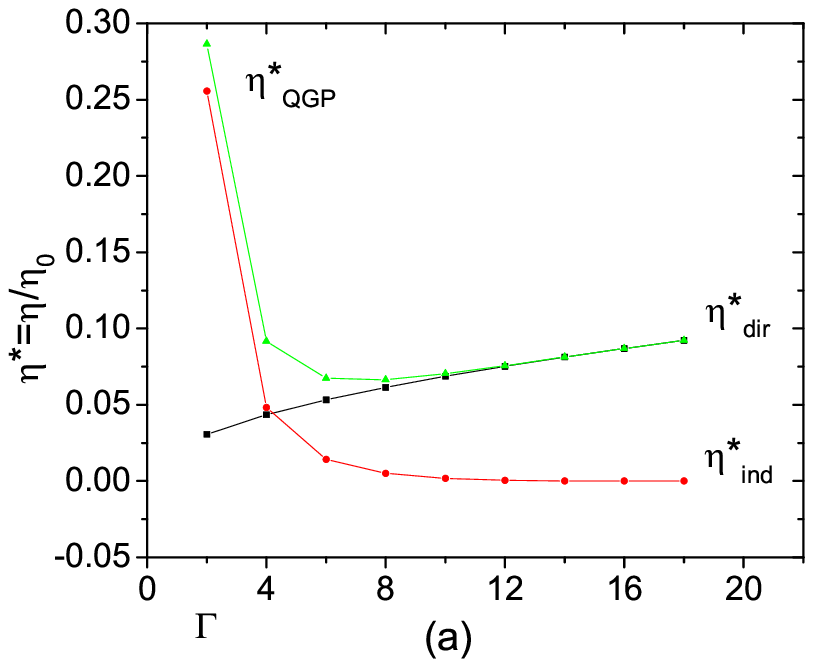}
\includegraphics[width=0.49\textwidth]{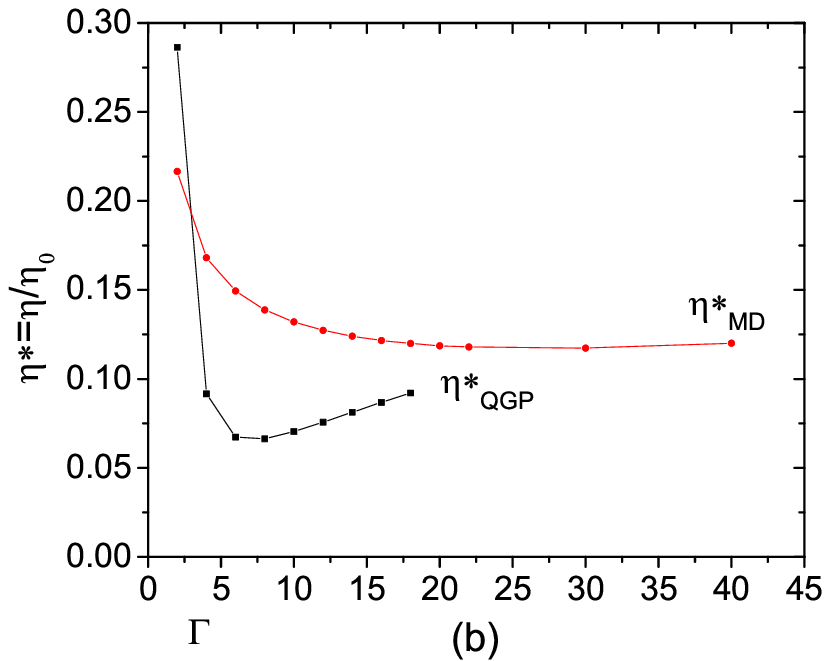}
\end{center}
\caption{The direct and indirect part of the
viscosity}\label{viscosity}
\end{figure}

\begin{figure}[!h]
\begin{center}
\includegraphics[width=0.49\textwidth]{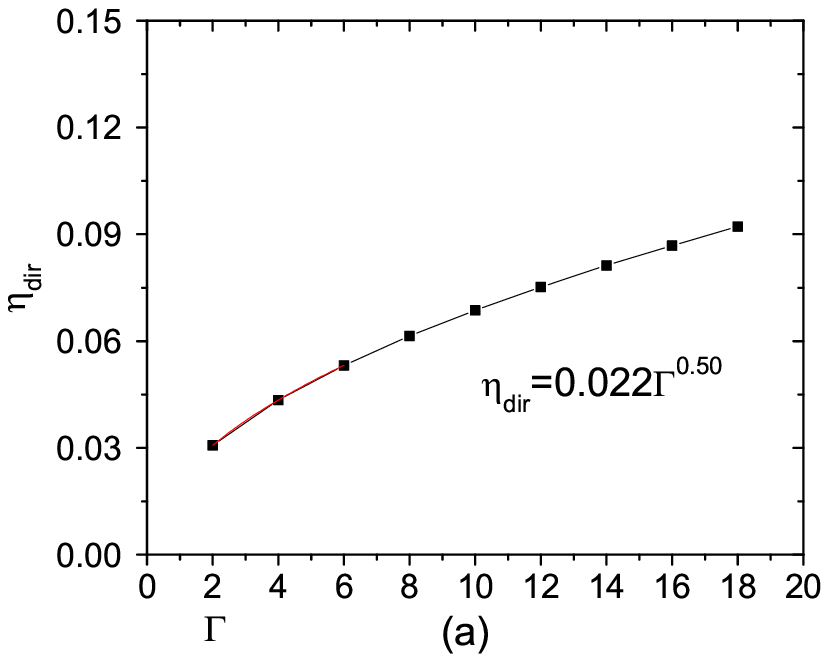}
\includegraphics[width=0.49\textwidth]{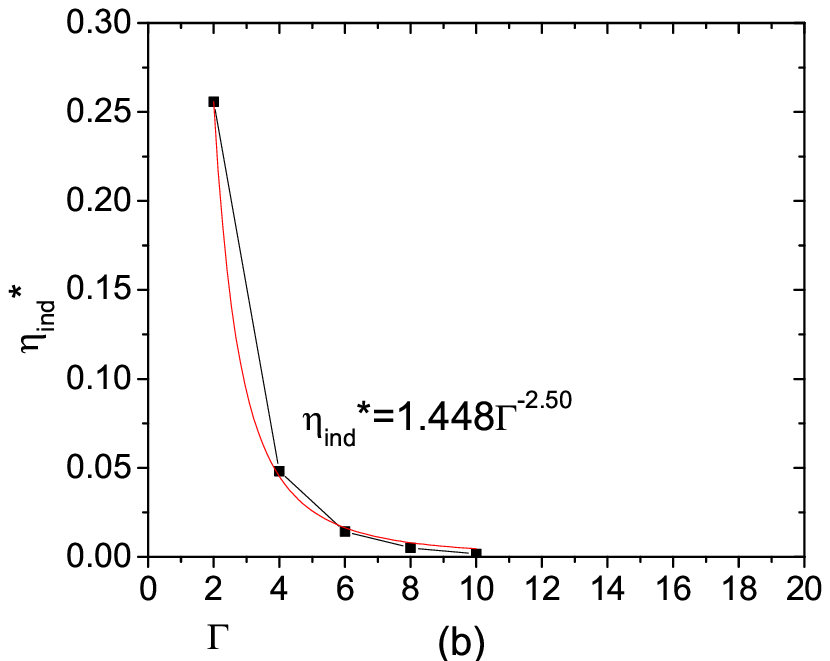}
\end{center}
\caption{The best fit of the direct and indirect part of the
viscosity}\label {viscosity_fit}
\end{figure}

Using the fitted monopole structure factor (\ref{FIT}) in
(\ref{eq014v}) we can numerically assess (\ref{eq013v}) for
different values of $\Gamma$.  Combining this result for the
indirect viscosity together with (\ref{eq012v}) for the direct
viscosity yield the colorless or sound viscosity $\eta_0$.  The
values of $\eta_0$ are displayed  in Table I, and shown in
Fig.~\ref{viscosity} (black).  The SU(2) molecular dynamics
simulations in~\cite{gelmanetal}  which are parameterized as

\begin{equation}
\eta^{*}_{MD}\simeq0.001\Gamma+\frac{0.242}{\Gamma^{0.3}}+\frac{0.072}{\Gamma^2}
\label{eq011v}
\end{equation}
are also displayed in Table I and shown in Fig.~\ref{viscosity}
(red) for comparison. The sound viscosity dips at about
$\Gamma\approx 8$ in our analytical estimate. To understand the
origin of the minimum, we display in Fig. \ref{viscosity_fit} the
scaling with $\Gamma$ of the direct or hydrodynamical and the
indirect part of the shear viscosity. The direct contribution to
the viscosity grows like $\Gamma^{1/2}$, the indirect contribution
drops like $1/\Gamma^{5/2}$.  The latter dominates at weak
coupling, while the former dominates at strong coupling. This is
indeed expected, since the direct part is the contribution from
the hydrodynamical part of the phase space, while the indirect
part is the contribution from the non-hydrodynamical or
single-particle part of  phase space.  The crossing is at
$\Gamma\approx 4$.

\begin{table}[!h]
\caption{Reduced shear viscosity. See text.} \label{tb1}

\begin{center}
\begin{tabular}{cccccccccccc}
\hline

$\Gamma        $ & $    2  $ & $    4  $ & $   6   $ & $    8  $ & $   10  $ & $   12  $ & $   14  $ & $   16  $ & $   18  $  \\
\hline
$\eta^{*}_{QGP}$ & $ 0.286 $ & $ 0.092 $ & $ 0.067 $ & $ 0.066 $ & $ 0.070 $ & $ 0.076 $ & $ 0.081 $ & $ 0.087 $ & $ 0.092 $  \\
$\eta_{MD}     $ & $ 0.217 $ & $ 0.168 $ & $ 0.168 $ & $ 0.139 $ & $ 0.132 $ & $ 0.127 $ & $ 0.124 $ & $ 0.122 $ & $ 0.120 $  \\
\hline

\end{tabular}
\end{center}
\end{table}

The reduced sound velocity $\eta_*$ is dimensionless. To restore
dimensionality and compare with expectations for an SU(2) colored
Coulomb plasma, we first note that the particle density is about
$3\times\,0.244\,T^3=0.732\,T^3$. There are 3 physical gluons,
each carrying black-body density. The corresponding Wigner-Seitz
radius is then $a_{WS}=({3}/{4\pi n})^{{1}/{3}}\approx
{0.688}/{T}$. The Coulomb coupling is $\Gamma\approx
1.453\,({g^2N_c}/{4\pi})$. Since the plasmon frequency is
$\omega_p^2={\kappa_D^2}/{m\beta}={n}g^2N_c/m$, we get
$\omega_p^2\simeq3.066\,T^2({g^2N_c}/{4\pi})$ with $m\simeq3T$.
The unit of viscosity  $\eta_0=nm\omega_pa_{WS}^2$ translates to
$1.822\,T^3({g^2}N_c/4\pi)^{{1}/{2}}$.  In these units, the
viscosity for the SU(2) cQGP dips at about $0.066$ which is
$\eta^*_{QGP}\approx 0.066\,\eta_0\approx
0.120\,T^3\,(g^2N_c/4\pi)^{1/2}$. Since the entropy in our case is
$\sigma=6\,(4\pi^2/90)T^3$, we have for the SU(2) ratio
$\eta/\sigma|_{SU(2)}=0.046\,(g^2N_c/4\pi)^{1/2}$. The minimum in
the viscosity occurs at $\Gamma=1.453\,({g^2N_c}/{4\pi})\approx
8$, so that $({g^2N_c}/{4\pi})^{1/2}\approx 2.347$. Thus, our
shear viscosity to entropy ratio is
$\eta/\sigma|_{SU(2)}\simeq0.107$. A rerun of these estimates for
SU(3) yields $\eta/\sigma|_{SU(3)}\simeq0.078$ which is lower than
the bound $\eta/\sigma=1/4\pi\simeq0.0795$ suggested from
holography.


\begin{figure}[!h]
\begin{center}
\includegraphics[width=0.49\textwidth]{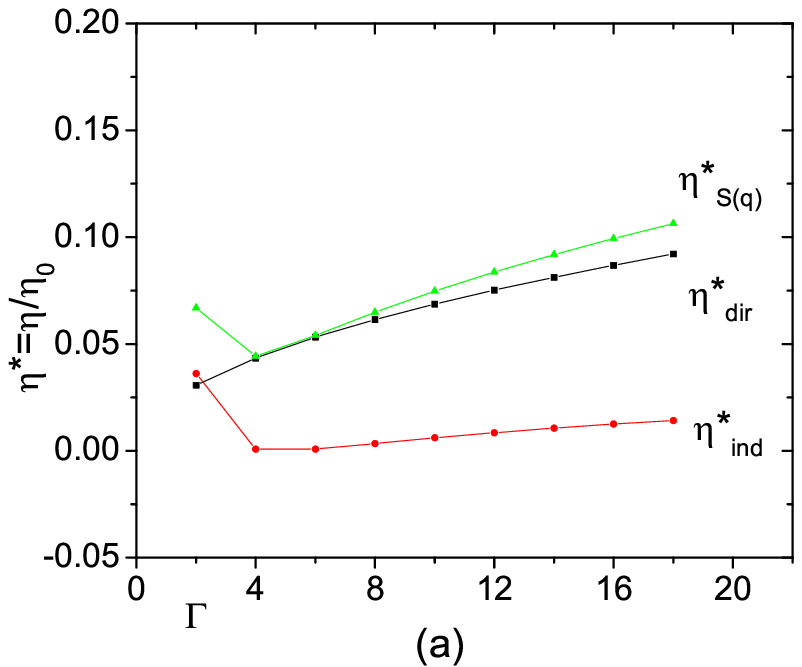}
\includegraphics[width=0.49\textwidth]{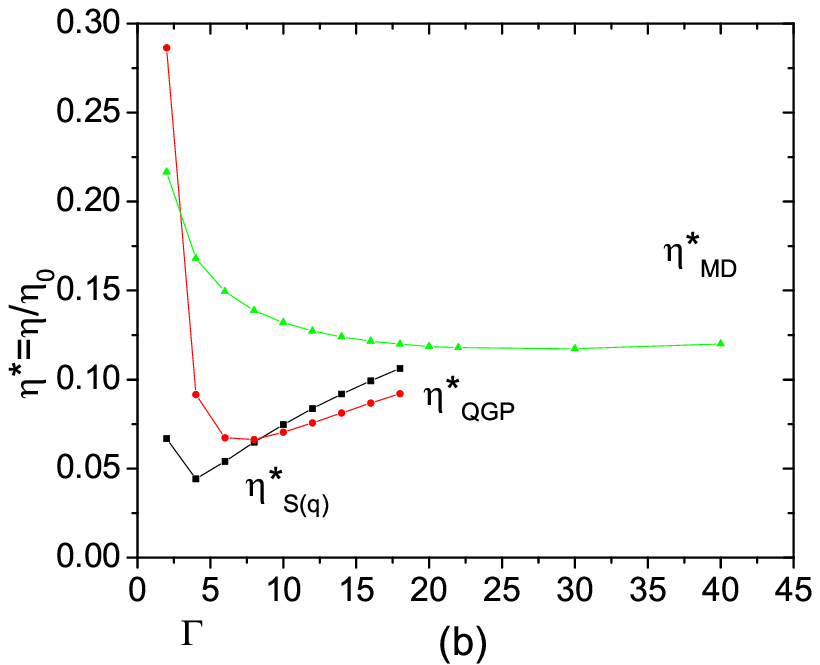}
\end{center}
\caption{Comparison with weak coupling. See text.}
\label{viscosity_approx}
\end{figure}

Finally, we show in Fig.~\ref{viscosity_approx} the shear
viscosity $\eta*_{S(q)}$ at low $\Gamma$ (a:green) and large
$\Gamma$ (b:black) assessed using the weak-coupling structure
factor $S(k)=k^2/(k^2+k_D^2)$.  The discrepancy is noticeable for
$\Gamma$ near the liquid point. The large discrepancy for small
values of $\Gamma$ reflects on the fact that the integrals in
(\ref{eq014v}) are infrared sensitive. The sensitivity is tamed by
our analytical structure factor and the simulations. We recall
that in weak coupling, the Landau viscosity $\eta_L$
is~\cite{ichimaru3}

\begin{equation}
\frac{\eta_L}{\eta^*}=\frac{5\sqrt{3\pi}}{18}\frac{1}{\Gamma^{5/2}}\frac 1{{\rm ln}(r_D/r_0)}
\label{LANDAU}
\end{equation}
which follows from a mean-field analysis of the kinetic equation with the
plasma dielectric constant set to 1. The logarithmic dependence in (\ref{LANDAU})
reflects on the infrared and ultraviolet  sensitivity of the mean-field approximation.
Typically $r_D=1/k_D$ and $r_0=(g^2C_2/4\pi)\beta$ which are the Debye length
and the the distance of closest approach. Thus

\begin{eqnarray}
\frac{\eta_L}{\eta^*}\approx\frac{5\sqrt{3\pi}}{27}\frac{1}{\Gamma^{5/2}}\frac 1{{\rm ln}(1/\Gamma)}
\label{LANDAU1}
\end{eqnarray}
or $\eta_L/\eta^*\approx 0.6/(\Gamma^{5/2}{\rm ln}(1/\Gamma))$ which is overall consistent
with our analysis.

The Landau or mean-field result  is smaller for the viscosity than the result
from perturbative QCD.  Indeed, the unscaled Landau viscosity (\ref{LANDAU1})
reads

\begin{equation}
\eta_L\approx \frac{10}{24}\frac{\sqrt{m}}{(\alpha_sC_2)^2\beta^{5/2}}\frac{1}{\alpha_s}
\label{LANDAU2}
\end{equation}
after restoring the viscosity unit $\eta^*=nm\omega_pa_{WS}^2$ and using
${\rm ln}(r_D/r_0)\approx 3{\rm ln}(1/\alpha_s)/2$ with $\alpha_s=g^2/4\pi$.
While our consituent gluons carry $m\approx \pi T$, in the mean field or weak
coupling we can set their masses to $m\approx gT$.  With this in mind, and setting $C_2=N_c=3$ in
(\ref{LANDAU2}) we obtain

\begin{equation}
\eta_L\approx \frac{5\sqrt{2}}{108\pi^{1/4}}\frac{T^3}{\alpha_s^{7/4}{\rm ln}(1/\alpha_s)}
\approx 0.05\frac{T^3}{\alpha_s^{7/4}{\rm ln}(1/\alpha_s)}
\label{LANDAU3}
\end{equation}
which is to be compared with the QCD weak coupling
result~\cite{heiselberg}

\begin{equation}
\eta_{QCD}\approx \frac{T^3}{\alpha_s^{2}{\rm ln}(1/\alpha_s)}
\label{LANDAU4}
\end{equation}
The mean-field result (\ref{LANDAU3}) is $\alpha_s^{1/4}\approx \sqrt{g}$ {\it smaller} in weak
coupling than the QCD perturbative result. The reason is the fact that in perturbative QCD the
viscosity is not only caused by collisions with the underlying parton constituents, but also
quantum recombinations and decays. These latter effects are absent in our classical QGP.

\section{Diffusion Constant}

\renewcommand{\theequation}{IX.\arabic{equation}}
\setcounter{equation}{0}

The calculation of the diffusion constant in the SU(2) plasma is
similar to that of the shear viscosity. The governing equation is
again (\ref{RES2}) with $\Sigma$ and ${\bf S}$  replaced by
$\Sigma_s$, ${\bf S}_s$. The label is short for single particle.
The difference between ${\bf S}$ and ${\bf S}_s$ is the
substitution of (\ref{FDIS}) by

\begin{equation}
f_s(\vec r\vec p \vec Q t)=\sqrt{N}\delta(\vec r-\vec
r_1(t))\delta(\vec p-\vec p_1(t))\delta(\vec Q-\vec Q_1(t))
\label{eq001t}
\end{equation}
The diffusion constant follows from the velocity auto-correlator

\begin{equation}
V_D(t)=\frac{1}{3}\langle\vec V(t)\cdot\vec V(0)\rangle
\label{eq003t}
\end{equation}
through

\begin{equation}
D=\int_{0}^{\infty}dt V_D(t)\label{eq002t}
\end{equation}
Solving (\ref{RES2}) using the method of one-Sonine polynomial
approximation as in~\cite{gould&mazenko} yields the Langevin-like
equation

\begin{equation}
\frac{dV_D(t)}{dt}=-\int_{0}^{t}dt'M(t')V_D(t-t') \label{eq004t}
\end{equation}
with the memory kernel tied to $\Sigma_{C0}^S$,

\begin{eqnarray}
&& n\,f_0(\vec p')\, \Sigma_{Cl}^S(t, \vec k, \vec p\vec p')
\nonumber \\
& & = -\frac{1}{\beta} \int d\vec p_1 d\vec p_2 \int \frac{d\vec
l}{(2\pi)^3} \vec l\cdot\nabla_{\vec p}\vec l\cdot\nabla_{\vec p'}
{\bf c}_{D1}(l) V_{\vec l} \nonumber \\
& & \times \Big( \frac{l}{2l+1} {\bf S}_{l-1}^S(t, \vec k-\vec
l,\vec p\vec p'){\bf S}_1(t, \vec l,\vec p_1\vec
p_2)+\frac{l+1}{2l+1} {\bf S}_{l+1}^S(t, \vec k-\vec l,\vec p\vec
p'){\bf S}_1(t, \vec l,\vec p_1\vec p_2)\Big) \nonumber \\
\label{eq005d}
\end{eqnarray}
and

\begin{eqnarray}
&& n\,f_0(\vec p')\, \Sigma_{C0}^S(t, \vec k=\vec 0, \vec p\vec
p')
\nonumber \\
& & = -\frac{1}{\beta} \int d\vec p_1 d\vec p_2 \int \frac{d\vec
l}{(2\pi)^3}\vec l\cdot\nabla_{\vec p}\vec l\cdot\nabla_{\vec p'}
{\bf c}_{D1}(l)V_{\vec l} {\bf S}_{1}^S(t,\vec l,\vec p\vec
p'){\bf S}_1(t, \vec l,\vec p_1\vec p_2) \label{eq006d}
\end{eqnarray}
therefore

\begin{equation}
M(t)=\frac{\beta}{3m}\int d\vec p d\vec p' \vec p\cdot\vec p'
\Sigma_{C0}^S(t,\vec k=\vec 0,\vec p\vec p')f_0(\vec p')
\label{eq005t}
\end{equation}
which clearly projects out the singlet color contribution. If we
introduce the dimensionless diffusion constant,
$D^{*}=D/w_pa_{WS}^2$,  then (\ref{eq002t}) together with
(\ref{eq004t}) yield

\begin{equation}
\frac{1}{D}=m\beta\int_{0}^{\infty}dtM(t) \rightarrow
\frac{1}{D^{*}}=3\Gamma\int_{0}^{\infty}w_p dt\frac{M(t)}{w_p^2}=
3\Gamma\int_{0}^{\infty} d\tau\bar{M}(\tau) \label{eq006t}
\end{equation}
Using similar steps as for the derivation of the viscosity, we can
unwind the self-energy kernel $\Sigma_s$ in (\ref{eq006t}) to give

\begin{equation}
\frac{1}{D^{*}}=-\Gamma\int\frac{d\vec l}{(2\pi
)^3}\int_{0}^{\infty}d\tau \vec l^2 {\bf c}_{D1}(l)V_{\vec l}
{\cal G}_{n1}^S(l,t)\,{\cal G}_{n1}(l,t) \label{eq007t}
\end{equation}
where we have used the same the half-renormalization method
discussed above for the viscosity.  The color integrations are
done by Legendre transforms. Here again, we separate the
time-dependent structure factors as ${\cal G}_{n1}(l,t)={\bf
S}_{01}(l)\bar{\cal G}_{n1}(l,t)$ and ${\bf
S}_{01}^S(l,t)=\bar{\cal G}_{n1}(l,t)$ in the free particle
approximation. Thus

\begin{equation}
\frac{1}{D^{*}}=\Gamma^{3/2}\Big(\frac{1}{3\pi}
\Big)^{\frac{1}{2}}\int_{0}^{\infty} dq q(1-{\bf S}_{01}(q))
\label{eq008t}
\end{equation}

\begin{figure}[!h]
\begin{center}
\includegraphics[width=0.55\textwidth]{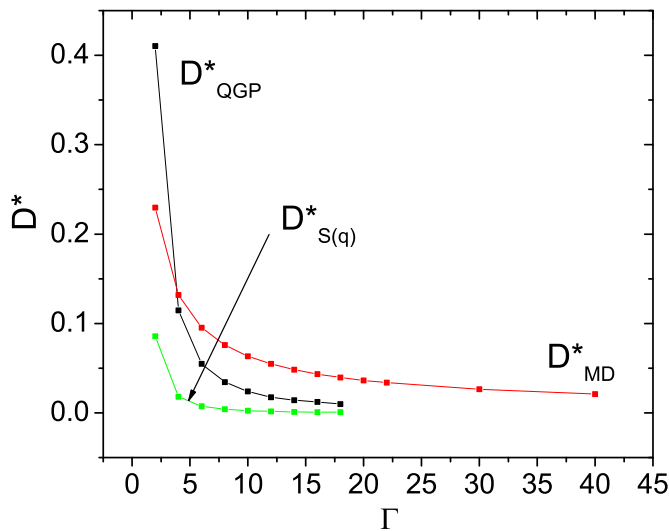}
\end{center}
\caption{Diffusion Constant (black, green) versus molecular dynamics
simulations (red). See text. } \label{diffusion}
\end{figure}

\begin{table}[!h]
\caption{Diffusion constant. See text.}
 \label{tb2}

\begin{center}
\begin{tabular}{cccccccccc}
\hline

$\Gamma       $ & $    2  $ & $   4   $ & $   6   $ & $    8  $ & $   10  $ & $   12  $ & $  14   $ & $   16  $ & $   18  $ \\
\hline
$D^{*}_{QGP}  $ & $ 0.410 $ & $ 0.115 $ & $ 0.055 $ & $ 0.034 $ & $ 0.024 $ & $ 0.017 $ & $ 0.014 $ & $ 0.012 $ & $ 0.010 $ \\
$D^{*}_{MD}   $ & $ 0.230 $ & $ 0.132 $ & $ 0.095 $ & $ 0.076 $ & $ 0.063 $ & $ 0.055 $ & $ 0.048 $ & $ 0.044 $ & $ 0.040 $ \\
\hline \label{DIFF}
\end{tabular}
\end{center}
\end{table}

The results following from (\ref{eq008t}) are displayed in
Table~\ref{DIFF} and in Fig.~\ref{diffusion} (black) from weak to
strong coupling. For comparison, we also show the the diffusion
constant measured using molecular dynamics simulations with an
SU(2) colored Coulomb plasma~\cite{gelmanetal}. The molecular
dynamics simulations are fitted to

\begin{equation}
D^{*}\simeq\frac{0.4}{\Gamma^{0.8}}
\end{equation}
For comparison, we also show the diffusion constant (\ref{eq008t})
assessed using the weak coupling or Debye structure factor $S(k)=k^2/(k^2+k_D^2)$
in Fig.~\ref{diffusion} (green).  The discrepancy between the analytical
results at small $\Gamma$ are similar to the ones we noted above
for the shear viscosity. In our correctly resummed structure factor
of Fig.~\ref{fluctuation}, the infrared behavior of the cQGP is controlled
in contrast to the simple Debye structure factor.

\begin{figure}[!h]
\begin{center}
\includegraphics[width=0.495\textwidth]{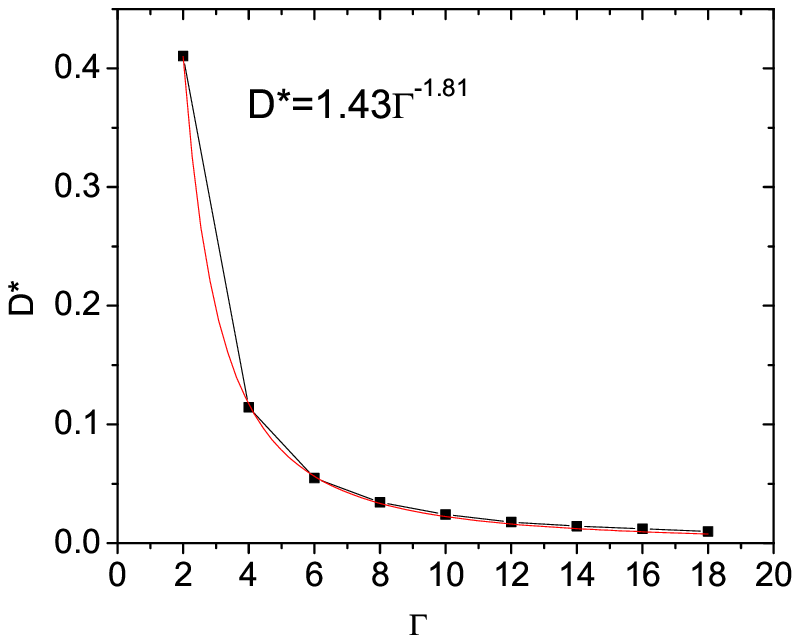}
\includegraphics[width=0.495\textwidth]{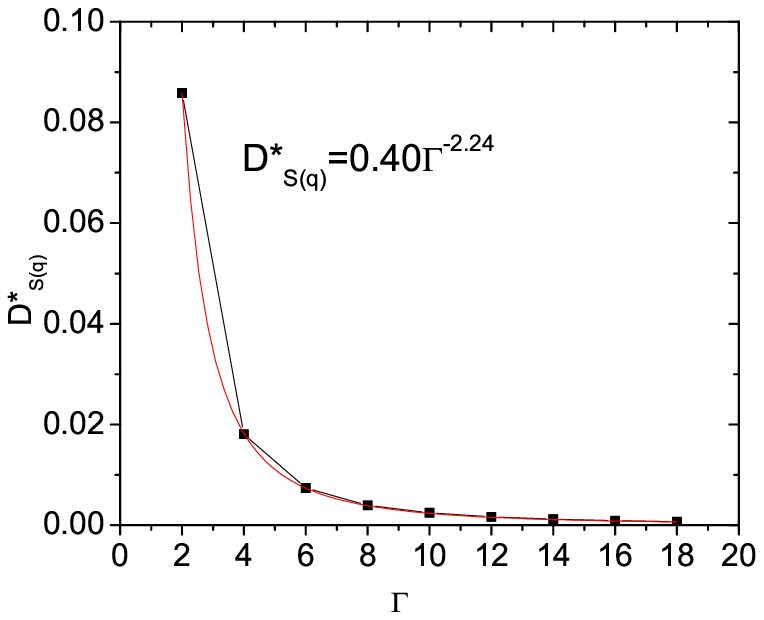}
\end{center}
\caption{Fit to the diffusion constant. See text.}
\label{diffusion_fit}
\end{figure}

 Finally, a comparison of (\ref{eq008t}) to (\ref{eq014v}) shows that
$1/D^*\approx 1/\lambda I_3$ which is seen to grow like
$\Gamma^{3/2}$. Thus $D^*$ drops like $1/\Gamma^{3/2}$
which is close to the numerically generated result fitted in
Fig.~\ref{diffusion_fit} (left). The weak coupling self-diffusion
coefficient scales as $1/\Gamma^{5/2}$ as shown in Fig.~\ref{diffusion_fit} (right).
 More importantly, the diffusion constant in the SU(2) colored Coulomb
plasma is caused solely by the non hydrodynamical modes or single
particle collisions in our analysis.
It does not survive at strong coupling where
most of the losses are caused by the collective sound and/or
plasmon modes. This result is in contrast with the shear viscosity we
discussed above, where the hydrodynamical modes level it off at large $\Gamma$.

\section{Conclusions}

We have provided a general framework for discussing
non-perturbative many-body dynamics in the colored SU(2) Coulomb
plasma introduced in \cite{SZ_newqgp}. The framework extends the
analysis developed intially for one-component Abelian plasmas to
the non-Abelian case. In the latter, the Liouville operator is
supplemented by a color precessing contribution that contributes
to the connected part of the self-energy kernel.

The many-body content of the SU(2) colored Coulomb plasma are best
captured by the Liouville equation in phase space in the form of
an eigenvalue-like equation. Standard projected perturbation
theory like analysis around the static phase space distributions
yield a resummed self energy kernel in closed form. Translational
space invariance and rigid color rotational invariance in phase
space simplifies the nature of the kernel.

In the hydrodynamical limit, the phase space projected equations
for the time-dependent and resummed structure factor displays both
transverse and longitudinal hydrodynamical modes. The shear
viscosity and longitudinal diffusion constant are expressed
explicitly in terms of the resummed self-energy kernel. The latter
is directly tied with the interacting part of the Liouville
operator in color space. We have shown that in the free streaming
approximation and half-renormalized Liouville operators, the
transport parameters are finite.

We have explicitly derived the shear viscosity and longitudinal
diffusion constant of the SU(2) colored Coulomb plasma in terms of
the monopole static structure factor and the for all values of the
classical Coulomb parameter $\Gamma=V/K$, the ratio of the
potential to kinetic energy per particle. The results compare
fairly with molecular dynamics simulations for SU(2).

The longitudinal diffusion constant is found to drop from weak to
strong coupling like $1/\Gamma^{3/2}$. The shear viscosity is
found to reach a minimum for $\Gamma$ of about 8. The large
increase at weak coupling is the result of the large mean free
paths and encoded in the direct or driving part of the connected
self-energy. The minimum at intermediate $\Gamma$ is tied with the
onset of hydrodynamics which reflects on the liquid nature of the
colored Coulomb plasma in this regime.

At larger values of $\Gamma$ an SU(2) crystal forms as reported
in~\cite{SZ_newqgp}. Our current analysis should be able to
account for the emergence of elasticities, with in particular an
elastic shear mode. This point will be pursued in a future
investigation. The many body analysis presented in this work treats the color
degrees of freedom as massive constituents with a finite mass and
a classical SU(2) color charge. The dynamical analysis is fully
non-classical. In a way, quantum mechanics is assumed to generate
the constituent degrees of freedom with their assigned parameters.
While this picture is supported by perturbation theory at very
weak coupling, its justification at strong coupling is by no means
established.

\begin{acknowledgements}
This work was supported in part by US DOE grants DE-FG02-88ER40388
and DE-FG03-97ER4014.
\end{acknowledgements}

\newpage

\appendix


\section{SU(2) color phase space}

\renewcommand{\theequation}{A.\arabic{equation}}
\setcounter{equation}{0}

A useful parametrization of the SU(2) color phase space is through the canonical variables
$Q^1,\pi^1$~\cite{johnson,litim&manuel}

\begin{equation}
Q^1=\cos\phi_1 \sqrt{J^2-\pi_1^2}, \quad  Q^2=\sin\phi_1
\sqrt{J^2-\pi_1^2}, \quad Q^3=\pi^1 \label{eq001aa}
\end{equation}
with $Q^2$ being a constraint variable fixed by
$J^2$ or the quadratic Casimir with $q_2=
\sum_{\alpha}^{N_c^2-1}{Q^{\alpha}Q^{\alpha}}$.
The conjugate set $Q^1, \pi^1$ obeys standard Poisson bracket.
The associated phase space measure is

\begin{equation}
dQ=c_R d\pi_1 d\phi_1 J dJ \delta(J^2-q_2) \label{eq002aa}
\end{equation}
where $c_R$ is a representation dependent constant.
A simpler parametrization of the phase space is to use

\begin{equation}
dQ=\sin{\theta}d\theta d\phi \label{eq005aa}
\end{equation}
with the normalizations $\int dQ=4\pi$,
$\sum_{\alpha}Q^{\alpha}Q^{\alpha} =1$ and $\int dQ \vec
Q\cdot\vec Q=4\pi$.  The SU(2) Casimir is then restored by inspection.


\section{Projection Method}

\renewcommand{\theequation}{B.\arabic{equation}}
\setcounter{equation}{0}

If we define the phase space density, $\delta f_l^m(\vec k\vec
p,t)$

\begin{equation}
\delta f_l^m(\vec k \vec p, t)=\sum_{i=1}^N e^{-i\vec k\cdot\vec
r_i(t)}\delta(\vec p-\vec p_i(t)){Y_l^m}(\vec Q_i)-nf_0(\vec p)
\delta_{l0}\delta_{m0}\delta_{\vec k\vec 0}Y_0^0 \label{eq001e}
\end{equation}
we can construct structure factor ${\bf S}_l(t,\vec k,\vec p\vec
p')$ for $l$th partial wave

\begin{eqnarray}
& & \frac{4\pi}{2l+1}\sum_m (\delta {f_l^m}^{*}(\vec k \vec p,
t)|\delta f_l^m(\vec k \vec p', 0)) \nonumber \\
& & \equiv \frac{4\pi}{2l+1}\sum_m\sum_{i,j}^N( e^{-i\vec k\cdot
(\vec r_i(t)-\vec r_j(t))}\delta(\vec p-\vec p_i(t))\delta(\vec
p'-\vec p_j) {Y_l^m}^*(\vec Q_i){Y_l^m}(\vec Q_j))
-n^2f_0(\vec p)f_0(\vec p') \nonumber \\
& & \equiv {\bf S}_l(t,\vec k,\vec p\vec p') \label{eq002e}
\end{eqnarray}
Here a scalar product $(A|B)$ is defined as $\langle
A^*B\rangle_{eq}$. We follow~\cite{mori,akcasu&duderstadt,baus}
and recast the formal Liouville equation (\ref{RES}) in the form
of a formal eigenvalue-like equation in phase space

\begin{equation}
{\bf S}_l(\vec k z ;\vec p \vec p')=(\delta {f_l^m}^{*}(\vec k\vec
p) |(z-\mathcal{L})^{-1}| \delta f_l^m(\vec k'\vec p'))
\label{eq003e}
\end{equation}
The color charge effect by partial waves is represented as $l,m$
in Eq. (\ref{eq003e}). If we introduce the projection operator

\begin{equation}
\mathcal{P}=4\pi\sum_{l,m,\vec k}\int d\vec p_1 d\vec p_2|\delta
f_l^m(\vec k,\vec p_1)\rangle {\bf S}_{0l}^{-1}(\vec k,\vec
p_1,\vec p_2)\langle\delta {f_l^m}^*(\vec k,\vec
p_2)|=1-\mathcal{Q} \label{eq004e}
\end{equation}
we can check that this projection operator satisfies
$\mathcal{P}^2=\mathcal{P}$

\begin{eqnarray}
\mathcal{P}^2 & =& 4\pi\sum_{l,m,\vec k}\sum_{l',m',\vec k'}\int
d\vec p_1 d\vec p_2d\vec p_1'd\vec p_2'|\delta f_l^m(\vec k,\vec
p_1)){\bf S}_{0l}^{-1}(\vec k,\vec p_1,\vec p_2) \nonumber \\
& \times & 4\pi (\delta {f_l^m}^*(\vec k,\vec p_2)|\delta
{f_{l'}^{m'}}(\vec k,\vec p_1')){\bf S}_{0l}^{-1}(\vec k,\vec
{p_1}',\vec {p_2}')(\delta {f_{l'}^{m'}}^*(\vec k,\vec
{p_2}')|=\mathcal{P} \label{eq005e}
\end{eqnarray}
because of the translational invariance in space and the
rotational invariance in color space,

\begin{equation}
4\pi(\delta {f_l^m}^*(\vec k,\vec p_2)|\delta {f_{l'}^{m'}} (\vec
k,\vec p_1'))\equiv\delta_{\vec k\vec k'}\delta_{ll'}
\delta_{mm'}{\bf S}_{0l}(\vec k,\vec p_2,\vec {p_1}')
\label{eq006s}
\end{equation}
The off-diagonal elemenets vanish in the equilibrium averaging due
to phase incoherence. Therefore, the projection operator in Eq.
(\ref{eq005e}) satisfies also $\mathcal{Q}^2=\mathcal{Q}$ and $
\mathcal{P}\mathcal{Q} =\mathcal{Q}\mathcal{P}=0$. If we define
$|F_l^m(\vec k\vec p;z))$ as $|F_l^m(\vec k\vec
p;z))=(z-\mathcal{L})^{-1}|\delta f_l^m(\vec k\vec p))$ from Eq.
(\ref{eq003e}), we have

\begin{equation}
\mathcal{P}(z-\mathcal{L})|F_l^m(\vec k\vec
p;z))=\mathcal{P}|\delta f_l^m(\vec k\vec p)) \label{eq007e}
\end{equation}
$\mathcal{P}$ in Eq. (\ref{eq005e}) is the operator which projects
phase space function of a multipartle state with $l'$th partial
wave into a single particle state of the same parial wave,
$|\delta f_{l'}^{m'}(\vec k\vec p))$,
$\mathcal{P}|g_{l'}^{m'}(\vec k\vec p))=|\delta f_{l'}^{m'}(\vec
k\vec p))$. Therefore $\mathcal{Q}|\delta f_l^m(\vec k\vec
p))=(1-\mathcal{P})|\delta f_l^m(\vec k\vec p))=0$. With these in
mind, we can modify the above equation further using
$\mathcal{P}+\mathcal{Q}=I$

\begin{eqnarray}
& & (\mathcal{P}z-\mathcal{P}\mathcal{L}\mathcal{P}-\mathcal{P}
\mathcal{L} \mathcal{Q}) |F_l^m(\vec k\vec p;z))
=\mathcal{P}|\delta f_l^m(\vec k\vec p)) \nonumber \\
& & (\mathcal{Q}z-\mathcal{Q}\mathcal{L}\mathcal{P}-\mathcal{Q}
\mathcal{L} \mathcal{Q}) |F_l^m(\vec k\vec p;z))=0 \label{eq008e}
\end{eqnarray}
From these equations, we can extract

\begin{equation}
z\mathcal{P}|F_l^m(\vec k\vec p;z))-\mathcal{P}
\mathcal{L}\mathcal{P}|F_l^m(\vec k\vec p;z))
-\mathcal{P}\mathcal{L}\mathcal{Q}(z-\mathcal{Q}\mathcal{L}
\mathcal{Q})^{-1}\mathcal{Q}\mathcal{L}\mathcal{P}|F_l^m(\vec
k\vec p;z))=\mathcal{P}|\delta f_l^m(\vec k\vec p)) \label{eq009e}
\end{equation}
By multiplying $(\delta f(\vec k\vec p)|$ we finally obtain,

\begin{equation}
z{\bf S}_l(\vec k z;\vec p \vec p')-\int d\vec p_1d\Sigma_l(\vec k
z;\vec p \vec p_1){\bf S}_l(\vec k z;\vec p_1 \vec p')={\bf
S}_l(\vec k 0;\vec p \vec p') \label{eq010e}
\end{equation}
where the memory function, or the evolution operator
$\Sigma_l(\vec k z;\vec p \vec p_1)$ is

\begin{equation}
\Sigma_l(\vec k z;\vec p \vec p')=\frac{4\pi}{2l+1}\sum_m \int
d\vec p_1(\delta {f_l^m}^*(\vec k,\vec p)|\mathcal{L}+{
\Psi}|\delta f_l^m(\vec k,\vec p_1)){\bf S}_{0l}^{-1}(\vec k,\vec
p_1,\vec p') \label{eq011e}
\end{equation}
with

\begin{equation}
{\Psi}=\mathcal{P}\mathcal{L}\mathcal{Q}(z-\mathcal{Q}
\mathcal{L}\mathcal{Q})^{-1} \mathcal{Q}\mathcal{L}\mathcal{P}
\label{eq012e}
\end{equation}
Since the Liouville operator $\mathcal{L}$ can be split into
$\mathcal{L}_0+\mathcal{L}_I+\mathcal{L}_Q$, Eq. (\ref{LIOU1}),
the evolution operator can also be split into four terms; the free
streaming term($\Sigma^0_l$), the self consistent
term($\Sigma^s_l$), the color charge term($\Sigma^Q$) and the
non-local collision term($\Sigma^c$).

\begin{eqnarray}
& & \Sigma_{0l}(\vec k z;\vec p \vec p')=\frac{\vec k\cdot\vec
p}{m}\delta(\vec p-\vec p') \nonumber \\
& & \Sigma_{Il}(\vec k z;\vec p \vec p')=-n\frac{\vec k\cdot\vec
p}{m} f_0(\vec p){\bf c}_{Dl}(\vec k) \nonumber \\
& & \Sigma_{Ql}(\vec k z;\vec p \vec p')=0 \nonumber \\
& & \Sigma_{Cl}(\vec k z;\vec p \vec p')=\frac{1}{n f_0(\vec p)}
\frac{4\pi}{2l+1}\sum_m (\delta {f_l^m}^*(\vec k\vec
p)\vert\mathcal{L}\mathcal{Q}
(z-\mathcal{Q}\mathcal{L}\mathcal{Q})^{-1} \mathcal{Q}\mathcal{L}
\vert\delta f_l^m(\vec k\vec p')) \label{eq013e}
\end{eqnarray}

\section{Collisional Color Contribution}

\renewcommand{\theequation}{C.\arabic{equation}}
\setcounter{equation}{0}

In this Appendix we detail the calculation that leads to a zero
contribution from the colored Liouville operator in the
collisional part of the self energy in the free streaming
approximation. A typical contribution to (\ref{ST4}) and
(\ref{ST6}) is

\begin{eqnarray}
&&L_Q(\vec q,\vec q_1)\,L^R_Q(\vec q',\vec q_2)\,{\bf S}(t,
\vec q, \vec q_2)\,{\bf S}(t, \vec q', \vec q_1)=\frac 1\beta
\nonumber \\
&&\times\bigg(V(\vec r-\vec r_1)\,\vec Q\times\vec{Q}_1\cdot
(\vec\nabla_Q-\vec\nabla_{Q_1})\bigg)\nonumber\\
&&\times\bigg({\bf c}'_D(\vec r'-\vec r_2, \vec Q'\cdot \vec
Q_2))\, \vec Q'\times\vec{Q}_2\cdot(\vec\nabla_{Q'}
-\vec\nabla_{Q_2})\bigg)\,\,{\bf S}(t, \vec q, \vec q_2)\,{\bf
S}(t, \vec q', \vec q_1) \label{E1}
\end{eqnarray}
which can be reduced to

\begin{eqnarray}
&&L_Q(\vec q,\vec q_1)\,L^R_Q(\vec q',\vec q_2)\,{\bf S}(t, \vec
q, \vec q_2)\,{\bf S}(t, \vec q', \vec q_1)=-\frac 1\beta\,V(\vec
r-\vec r_1)\,{\bf c}'_D(\vec r'-\vec r_2, \vec Q'\cdot \vec
Q_2))\, \nonumber \\
&&\times\Bigg({\bf S}'(\vec Q\cdot \vec Q_2){\bf S}'(\vec Q'\cdot
\vec Q_1) (\vec Q_1\times \vec Q_2)\cdot \vec Q\,(\vec Q_1\times
\vec Q_2)\cdot \vec Q'\nonumber\\
&&\,\,\,\,\,\times {\bf S}''(\vec Q\cdot \vec Q_2){\bf S}(\vec Q'
\cdot \vec Q_1) (\vec Q_1\times \vec Q_2)\cdot \vec Q\,(\vec Q
\times \vec Q')\cdot \vec Q_2\nonumber\\
&&\,\,\,\,\,\times {\bf S}(\vec Q\cdot \vec Q_2){\bf S}''(\vec Q'
\cdot \vec Q_1) (\vec Q\times \vec Q')\cdot \vec Q_1\,(\vec Q_1
\times \vec Q_2)\cdot \vec Q'\nonumber\\
&&\,\,\,\,\,\times {\bf S}'(\vec Q\cdot \vec Q_2){\bf S}'(\vec
Q'\cdot \vec Q_1) (\vec Q\times \vec Q')\cdot \vec Q_1\,(\vec
Q\times \vec Q')\cdot \vec Q_2\Bigg) \label{E2}
\end{eqnarray}
The derivatives on ${\bf c}_D$ and ${\bf S}$ are on their color
argument.  We note that (\ref{E2}) contribute to the collisional
part of the self energy in (\ref{eq006s}) after the integration
over $Q_1$ and $Q_2$, which is then zero. This is expected.
Indeed, the colored Liouville operator is a 3-body force that
requires 3 distinct color charges to not vanish.  While (\ref{E2})
contributer to the unintegrated collisional operator, it does not
in the integrated one which is the self-energy on the 2point
function. It does contribute in the Liouville hierarchy in the
3-body structure factors and higher.

\section{Hydrodynamical subspace}

\renewcommand{\theequation}{D.\arabic{equation}}
\setcounter{equation}{0}

The projection method onto the hydrodynamical subspace has been discussed by
many~\cite{foster&martin, foster, baus}. This consists in dialing the projector in (\ref{K1})
onto the hydrodynamical modes. We choose Hermite polynomials as a basis set with
the Maxwell-Boltzman distribution $f_0(\vec{p})$ as a Gaussian weight function.  The
Hermite polynomials are the the generalized ones in 3D~\cite{grad}. Specifically

\begin{eqnarray}
& & H_{1(n)}(\vec p)=1 \quad\quad H_{2(l)}(\vec p)=p_z \quad\quad
H_{3(\epsilon)}(\vec p)=\frac{1}{\sqrt{6}}(p^2-3) \nonumber \\
& & H_{4(t_1)}(\vec p)=p_x \quad\quad H_{5(t_2)}(\vec p)=p_y
\label{eq001h}
\end{eqnarray}
These polynomials are orthonormal for the inner product

\begin{eqnarray}
& & \langle m|n\rangle=\int d\vec p a_mH_m(\vec p)a_nH_n(\vec
p) n f_0(\vec p)=\delta_{mn} \nonumber \\
& & \langle m|F(k,t)|n\rangle=\int d\vec pd\vec p' a_mH_m(\vec
p)F(k,t;\vec p\vec p')a_n H_n(\vec p')n f_0(\vec p')
\label{eq002h}
\end{eqnarray}
Here $a_m$ and $a_n$ set the normalizations. We chose the
longitudinal momentum direction along $\vec{k}$ in Fourier space,
$\langle l|=a_m\hat{k}\cdot\vec p$. The transverse directional is
chosen orthogonal to ${\vec k}$, $\langle
t|=a_m'\vec \epsilon\cdot\vec p$ with a unit vector satisfying
$\vec \epsilon^2=1$ and $\vec \epsilon\cdot\hat{k}=0$.

The hydrodynamical projection operators $\mathcal{P}_H$ restricted
to the five states (\ref{eq001h}) are

\begin{equation}
\mathcal{P}_H=\sum_i^{5}|i\rangle\langle i| \quad\quad
\mathcal{Q}_H=1-\mathcal{P}_H=1-\sum_i^{5}|i\rangle\langle i|
\label{eq003h}
\end{equation}
While in general these 5 statesare enough to characterize the
hydrodynamical modes in the SU(2) phase space, we need
additional states to work out the shear viscosity as it involves
in general correlations in the stress tensor through the Kubo
relation~\cite{hansen&mcdonald}. For that we need additionally,

\begin{equation}
H_6(\vec p)=p_xp_y \quad\quad H_7(\vec p)=p_xp_z \quad\quad
H_8(\vec p)=p_yp_z \label{eq004h}
\end{equation}
With the definition of $G_{ij}(\vec kz)=\langle i \vert {\bf
S}(\vec kz;\vec p \vec p')(nf_0(\vec p))^{-1} \vert j\rangle$ we
can rewrite (\ref{K1}) as

\begin{equation}
\Big(z-\sum_k\langle i \vert \Omega(\vec kz;\vec p \vec p') \vert
k\rangle\Big)G_{kj}(\vec kz)=G_{ij}(\vec k0) \label{eq005h}
\end{equation}
where $i,j$ are short for: $n$(density), $\epsilon$(energy), $l$(longitudinal
momentum) and $t$(transverse momentum).


{

}

\begin{thebibliography}{xx}

\bibitem{SZ_newqgp}
E.~V.~Shuryak and I.~Zahed,
Phys.\ Rev.\ C {\bf 70}, 021901 (2004) \\ 
E.~V.~Shuryak and I.~Zahed,
Phys.\ Rev.\ D {\bf 70}, 054507 (2004) 


\bibitem{hydro}
D.~Teaney, J.~Lauret and E.~V.~Shuryak, Phys.\ Rev.\ Lett.\  {\bf
86}, 4783 (2001) \\ 
D.~Teaney, J.~Lauret and E.~V.~Shuryak, {\tt nucl-th/0110037} \\
P. F. Kolb, P.Huovinen, U. Heinz, H. Heiselberg,
{ Phys. Lett.} {\bf B500} (2001)  232. \\ 
P.~F.~Kolb and U.~Heinz, {\tt nucl-th/0305084}

\bibitem{gelmanetal} B. A. Gelman, E. V. Shuryak and I. Zahed, Phys. Rev. C \textbf{74}, 044908 (2006)
\bibitem{cho&zahed} S. Cho and I. Zahed, Phys. Rev. C \textbf{79} 044911 (2009)
\bibitem{cho&zahed2} S. Cho and I. Zahed, Phys. Rev. C \textbf{80} 014906 (2009)
\bibitem{dusling&zahed} K. Dusling and I. Zahed, {\tt arXiv:0904.0169}
\bibitem{gelmanetal2} B. A. Gelman, E. V. Shuryak and I. Zahed, Phys. Rev. C \textbf{74}, 044909 (2006)

\bibitem{wong} S. K. Wong, Nuovo Cimento A \textbf{65}, 689 (1970)
\bibitem{cho&zahed3} S. Cho and I. Zahed, {\tt arXiv:0909.4725}

\bibitem{foster&martin} D. Foster and P. C. Martin, Phys. Rev. A \textbf{2}, 1575 (1970)
\bibitem{foster} D. Foster, Phys. Rev. A \textbf{9}, 943 (1974)
\bibitem{mazenko1} G. F. Mazenko, Phys. Rev. A. \textbf{7}, 209 (1973)
\bibitem{mazenko3} G. F. Mazenko, Phys. Rev. A. \textbf{9}, 360 (1974)
\bibitem{wallenborn&baus} J. Wallenborn and M. Baus, Phys. Rev, A \textbf{18}, 1737 (1978)
\bibitem{baus} M. Baus, Physica A \textbf{79}, 377 (1975)

\bibitem{cho&zahed5} S. Cho and I. Zahed, {\tt arXiv:0910.1548}
\bibitem{gould&mazenko} H. Gould and G. F. Mazenko, Phys. Rev. A \textbf{15}, 1274 (1977)

\bibitem{ichimaru3} S. Ichimaru, \textit{Statistical Plasma Physics Vol I:Basic Principles} (Westview Press, 2004)
\bibitem{heiselberg} H. Heiselberg, Phys. Rev. D \textbf{49}, 4739 (1994)



\bibitem{johnson} K. Johnson, Annals Phys. \textbf{192}, 101 (1989)
\bibitem{litim&manuel} D. F. Litim and C. Manuel, Phys. Rept. \textbf{364}, 451 (2002)
\bibitem{mori} H. Mori, Prog. Theor. Phys. \textbf{33}, 423 (1965)
\bibitem{akcasu&duderstadt} A. Z. Akcasu and J. J. Duderstadt, Phys. Rev. \textbf{188}, 479 (1969)

\bibitem{wallenborn&baus2} J. Wallenborn and M. Baus, J. Stat. Phys. \textbf{16}, 91 (1977)
\bibitem{grad} H. Grad, Comm. Pure Appl. Maths, \textbf{2}, 331 (1949)
\bibitem{hansen&mcdonald} J. P. Hansen and I. R. McDonald, \textit{Theory Of Simple Liquids, 3rd ed.} (Academic Press, 2006)






\end{thebibliography}
\end{document}